\documentclass[11pt]{article}
\pdfoutput=1
\usepackage{graphicx}
\usepackage{epsfig}
\usepackage{amsfonts}
\usepackage{amscd}
\usepackage{latexsym}
\usepackage{amsmath,amssymb}
\usepackage{verbatim}
\usepackage{setspace}
\usepackage{subfig}
\usepackage{color}
\usepackage{fancyhdr}
\usepackage{cite}
\usepackage{hyperref}
\usepackage{tensor}
\usepackage{xfrac}
\usepackage[textheight=9in, textwidth=6.5in, letterpaper]{geometry}
\usepackage{xcolor}
\usepackage{physics}
\usepackage{tcolorbox}
\usepackage{soul}
\usepackage{ulem}
\usepackage{float}
\usepackage{tikz}

\newcommand{\dA}{\dot{A}}
\newcommand{\dB}{\dot{B}}
\newcommand{\dC}{\dot{C}}
\newcommand{\dD}{\dot{D}}

\numberwithin{equation}{section}

{\cfoot{\thepage}}
\begin{document}
\begin{titlepage}
\unitlength = 1mm~\\
\vskip 3cm
\begin{center}
{\LARGE{\textsc{Black Holes in Klein Space}}}

\vspace{0.8cm}
Erin Crawley{$^{*}$}, Alfredo Guevara{$^{*\dagger\S}$}, Noah Miller{$^{*}$}, and Andrew Strominger{$^{*}$}\\
\vspace{1cm}

{$^*$\it  Center for the Fundamental Laws of Nature, Harvard University,\\
Cambridge, MA 02138, USA\\
$^\dagger$\it Black Hole Initiative, Harvard University, Cambridge, MA 02138, USA\\
$^\S$\it Society of Fellows, Harvard University, Cambridge, MA 02138, USA}

\vspace{0.8cm}

\begin{abstract}
The analytic continuation of the general signature $(1,3)$  Lorentzian Kerr-Taub-NUT black holes to signature $(2,2)$ Kleinian black holes is studied. Their global structure is characterized by a toric Penrose diagram resembling their Lorentzian counterparts.  Kleinian black holes are found to be self-dual when their mass and NUT charge are equal for any value of the Kerr rotation parameter $a$. Remarkably, it is shown that the rotation $a$ can be eliminated by a large diffeomorphism; this result also holds in Euclidean signature. The continuation from Lorentzian to Kleinian signature is naturally induced by the analytic continuation of the S-matrix. Indeed, we show that the geometry of linearized black holes, including Kerr-Taub-NUT, is captured by $(2,2)$ three-point scattering amplitudes of a graviton and a massive spinning particle.  This stands in sharp contrast to their Lorentzian counterparts for which the latter vanishes kinematically, and enables a direct link to the S-matrix.

\end{abstract}

\end{center}

\end{titlepage}

\tableofcontents

\section{Introduction}

Analytic continuation from $(1,3)$ Lorentzian to $(0,4)$ Euclidean signature has been an indispensable tool over the last century, leading  to a wealth of insights  into thermodynamics, quantum mechanics, quantum field theory and quantum gravity. More recently, continuation from $(1,3)$ Lorentzian to $(2,2)$ Kleinian signature has emerged as useful tool in quantum field theory and especially quantum gravity for several interrelated reasons:

\begin{itemize}
\item Recent studies of scattering  amplitudes using on-shell methods and spinor-helicity variables have led to expressions which are significantly simpler than  those arising in the traditional Euclidean-based Feynman diagramatics, see \cite{Elvang:2013cua,Taylor:2017sph} for an introduction. Spinor variables in conjugate representations are often explicitly or implicitly treated as independent, especially in the context of ``holomorphic'' soft limits which control much of their structure \cite{Cachazo:2014fwa,Cachazo:2014dia,He:2014bga,Casali:2014xpa,Elvang:2016qvq}. Doing so is equivalent to working in $(2,2)$ signature where the Lorentz group continues to two copies, self-dual and anti-self-dual,  of $SL(2,\mathbb{R})$. 
\item Much of the structure of amplitudes is controlled by the on-shell three-point function \cite{Britto:2005fq,Cachazo:2004kj,Arkani-Hamed:2008bsc,arkani2016grassmannian,Benincasa:2007xk}. However, nothing can be on-shell in Euclidean space (except tachyons!) while in Minkowski space kinematics set massless three-point amplitudes to zero. One is therefore forced to characterize massless scattering by significantly more complicated four-point scattering. In contrast, massless three-point amplitudes are generically nonzero in Klein space and usefully characterize the theory. 
\item  In quantum gravity, one seeks a holographic dual which lives on the conformal boundary of asymptotically flat spacetime \cite{ashtekar1987asymptotic,Strominger:2017zoo}. The conformal boundary of Euclidean space is a point,  whereas that of Klein space includes a Lorentzian torus \cite{Atanasov:2021celesttorus}. This ``celestial torus'' both provides a natural home for and constrains the nature of the sought-for holographic dual of 4D quantum gravity \cite{stromingerwalgebra, sharma2021ambidextrous,guevara2021celestial,adamo2021celestial,fan2019soft,Pate:2019lpp,Banerjee2020,puhm2020conformally,ebert2021descendants,atanasov2021conformal,Crawley:2021ivb,Himwich:2021dau,guevara2021holographic, Jiang:2021ovh, Adamo:2021lrv,ball2021perturbatively, mago2021deformed}.
\item Quantum self-dual gravity is  an interesting toy model for the study of more realistic 4D quantum theories of gravity \cite{OOGURI1991469,PhysRevD.54.7628,BERN1999423,Krasnov_2017}. As emphasized long ago, the self-duality condition can only be non-trivially imposed in Klein space, where it reduces to the quantization of geometries with $SL(2,\mathbb{R})$ holonomy and provides  connections to twistor theory \cite{Penrose:1976jq} and ultrahyperbolic geometries \cite{KO198151,flathspaces,EGUCHI1978249}.

\end{itemize}

Despite these developments there has been relatively little scrutiny of the global structure of Klein space. Conformal compactification was studied in \cite{Mason:2005qu} and the global structure of flat Klein space characterized by a toric Penrose diagram (defined below) in \cite{Atanasov:2021celesttorus}. In this paper we take steps towards a more global understanding  of Kleinian geometry by studying the Kleinian versions of the general stationary Kerr-Taub-NUT black hole spacetimes.  

Our most interesting results concern the case for which the mass and NUT charge are equal, which is self-dual for {\it any} value of the Kerr rotation parameter $a$. This makes them relevant to the the subject of quantum self-dual gravity. These have a simple toric Penrose diagram resembling the Penrose diagrams of their Lorentzian predecessors. We find however, that the rotation $a$ can be removed by a large diffeomorphism! This perhaps can be seen as the origin of  the many magical properties of the Kerr geometry. 
New Kerr coordinates are explicitly constructed for which the large diffeomorphism is a simple linear shift.  A  further  interesting property of Kleinian black holes is that they are in a sense on-shell.  Long ago \cite{PhysRevD.7.2317} Duff constructed the Lorentzian Schwarzschild geometry from {\it  off-shell} Feynman diagrams of gravitons with a massive source. In Klein space however, we show that black holes are described by radiative degrees of freedom captured by the {\it on-shell} three-point amplitude!
These simplifications bode well  for the potential utility of Kleinian black holes and their analysis using modern scattering methods.

This paper is organized as follows. 
In section 2, we will explain how one can analytically continue $(1,3)$ black hole metrics into $(2,2)$ signature. We will then narrow our focus to the self-dual  Taub-NUT metric, studying its global structure and drawing its toric Penrose diagram. We will also show that any angular momentum added to a self-dual Taub-NUT black hole  can be removed by a large diffeomorphism.

In section 3, we will show that stationary metrics in $(2,2)$ signature can be fully determined at linearized order from three-point graviton emission scattering amplitudes. This is in contrast to $(1,3)$ signature, where only the radiative part of the metric can be obtained. We will explicitly verify this for Kerr-Taub-NUT, and in doing so we land precisely on the new  coordinates from section \ref{sec:sec2} in which the rotation-shifting diffeomorphism is a simple shift of a single coordinate.\footnote{Indeed we found the new coordinates by working backwards from the amplitudes computation.}

In section 4, we will briefly discuss Euclidean Taub-NUT metrics and explain how some of our results about Kleinian black holes extend to Euclidean black holes. In particular, we will show that the Euclidean self-dual Kerr-Taub-NUT metric is diffeomorphic to the Euclidean self-dual Taub-NUT metric and explain how this result relates to previous literature. Moreover we will discuss a sense in which a generic non-self-dual Kerr-Taub-NUT geometry can be viewed as a relatively displaced superposition of self-dual and anti-self-dual non-rotating solutions. 

In section 5, we will end with a brief summary and discussion. 

In the appendices, we will outline the symmetries of the Kleinian Taub-NUT metric in more detail; construct via analytic continuation the Kleinian Schwarzschild and $M=0$ Taub-NUT metrics and their toric Penrose diagrams; compute the modular parameter of the Taub-NUT torus; outline the connection between radiative spacetimes and scattering amplitudes in Lorentzian signature; and finally compute the Kleinian Taub-NUT curvature tensor in terms of scattering amplitudes.

\section{Properties of Kleinian Black Holes}\label{sec:sec2} 

\subsection{Kleinian Taub-NUT }\label{sec:22TN}
In this subsection we construct the non-rotating\footnote{By ``non-rotating'' here and hereafter we simply mean we set the Kerr parameter  $a=\frac{J}{M}=0$;  depending on one's definitions rotation may also be associated with the NUT parameter $N$ introduced below.} Kleinian Taub-NUT solution by analytic continuation from Lorentzian signature. 
The Lorentzian Taub-NUT spacetime is a stationary, axisymmetric solution of Einstein's equations given by \cite{taub1951empty, nut1963empty}
\begin{equation}\label{Ltn}
    ds_{\textrm{TN, L}}^2 = -f_{\textrm{L}}(r)\left(dt - 2N \cos\theta d\phi\right)^2 +\frac{dr^2}{f_{\textrm{L}}(r)}+(r^2 + N^2)(d\theta^2 + \sin^2\theta d\phi^2)
\end{equation}
with
\begin{gather}
    f_{\textrm{L}}(r) = \frac{r^2 - 2 M r - N^2}{r^2 + N^2} = \frac{(r-r_+)(r-r_-)}{r^2 + N^2},\\
    r_\pm = M \pm \sqrt{ M^2 + N^2}.
\end{gather}
The metric is a generalization of the Schwarzschild solution, obtained by introducing a NUT charge  $N$, alongside the usual ADM mass $M$. Mass and NUT charge can be understood as being ``dual'' to each other in an analogy to electric and magnetic charge. When $N= 0$, the spacetime reduces to the Schwarzschild solution, whereas the $M = 0 $ limit is singular. For all $M$ and $N$, the metric has a $\mathfrak{su}(2) \times \mathfrak{u}(1)$ group of Killing symmetries.

The Lorentzian signature Taub-NUT metric \eqref{Ltn} suffers from well-known pathologies, as the presence of NUT charge can induce string-like singularities \cite{misner1963flatter, bonnor1969new}. In Euclidean signature, $M$ and $N$ have to be constrained (satisfying $M = \pm N$ or $M = \pm \tfrac{5}{4} N$) in order to avoid conical singularities \cite{HAWKING197781,Sorkin:1983ns, Gross:1983hb,gibbons1993gravitational}. We will see that in Kleinian signature neither of these issues arise.

There are many possible ways to analytically continue this metric from $(1,3)$ signature to $(2,2)$ signature. We will use the complex rotations
\begin{align}
    t &\to i t \label{eq:ttoit} \\
    \theta &\to i \theta \label{eq:thetatoitheta}\\
    N &\to i N.\label{dsc}
\end{align}
The fact that $N$ should be rotated is a consequence of the fact that $N$ is odd under time reversal, as sending $N \to - N$ is identical to sending $t \to -t$. Any rotation of $t$ will therefore be accompanied by a rotation of $N$ in order to retain the reality of the metric. (Likewise, when we later introduce the Kerr rotation parameter $a$, it too will  be rotated.) We will see in section \ref{sec:amplitudes} that the  resulting Kleinian metric  has a natural interpretation in terms of scattering amplitudes. The analytically continued Kleinian Taub-NUT metric is 
\begin{equation}
    ds^2_{\textrm{TN}} = f(r) (dt - 2 N \cosh \theta d \phi)^2 + \frac{dr^2}{f(r)} - (r^2 - N^2)( d \theta^2 + \sinh^2 \theta d \phi^2 )\label{eq_22TaubNUTgeneric},
\end{equation}
where
\begin{gather}
    f(r) = \frac{r^2 - 2 M r + N^2}{r^2 - N^2} = \frac{(r-r_+)(r-r_-)}{r^2 - N^2}\\
    r_\pm = M \pm \sqrt{ M^2 - N^2}.
\end{gather}
In order for the metric to be smooth on the surfaces $\theta = 0$ and $r = r_+$, in the generic case of $M \neq \pm N$ we must adopt the conditions
\begin{align}
    (t, \phi) &\sim (t - 4 \pi N, \phi + 2 \pi) \,\, \textrm{from}\,\, \theta=0\,,\label{eq_22TN_MnotN_period1}\\
    (t, \phi) &\sim (t + 4 \pi \frac{r_+^2 - N^2}{r_+ - r_-} , \phi)\,\,\textrm{from}\,\, r=r_+\,.\label{eq_22TN_MnotN_period2}
\end{align}
In the $M = \pm N$ case, $f(r)$ reduces to $f(r)=\frac{r+M}{r-M}$ and the periodicity conditions must be re-evaluated from scratch. They are found to be
\begin{align}
    (t, \phi) &\sim (t - 4 \pi N, \phi + 2 \pi)\label{eq_22TN_MisN_period1} \,\,\text{from}\,\,\theta=0, \\
    (t, \phi) &\sim (t + 4 \pi N, \phi + 2 \pi) \,\,\text{from}\,\,r=M\label{eq_22TN_MisN_period2}.
\end{align}
Note that the first of these cycles degenerates at $\theta = 0$, so in order to have a smooth metric after analytic continuation, we will impose the condition
\begin{equation}
    \theta \geq 0\,.\label{eq:thgeq0}
\end{equation}
The situation at $r = r_+$ is more involved; in the next section we will find that in the $M=\pm N$ case, there is a horizon at $r = r_+$. The spacetime \eqref{eq_22TaubNUTgeneric} has a four-dimensional Killing symmetry group of $\mathfrak{sl}(2, \mathbb{R}) \times \mathfrak{u}(1)$, as discussed in Appendix \ref{sec:app_TNsymmetries}. 

For generic $(M,N)$, the Kretschmann scalar is
\begin{equation}
 R_{\mu \nu \rho \sigma} R^{\mu \nu \rho \sigma} = 24 \left(\frac{(M - N)^2}{(N - r)^6} + \frac{(M + N)^2}{(N + r)^6}\right) \label{eq_22TNKretsch}
\end{equation}
which signals physical curvature singularities at $r= \pm N$ in the generic case $M \neq \pm N$. However, when $M = \pm N$, there will only be a singularity at $r = \mp N$. Finally, the metric \eqref{eq_22TaubNUTgeneric} is self-dual when $M = N$, satisfying
\begin{equation}
    R_{\mu \nu \rho \sigma} = \frac{1}{2} \varepsilon_{\mu \nu\alpha \beta} R^{\alpha \beta}_{\;\;\;\; \rho \sigma} \label{eq:rsdep}
\end{equation}
and anti-self-dual when $M = - N$, satisfying
\begin{equation}
    R_{\mu \nu \rho \sigma} = -\frac{1}{2} \varepsilon_{\mu \nu\alpha \beta} R^{\alpha \beta}_{\;\;\;\; \rho \sigma}  \label{eq:rsdep2}
\end{equation}
where we use the convention $\varepsilon_{t r \theta \phi} = \sqrt{|g|}$. Hence these are half-flat black holes with  holonomy group $\mathfrak{sl}(2, \mathbb{R}) $, a Kleinian analog of the $\mathfrak{su}(2) $  holonomy Euclidean Taub-NUT spaces. 

In general under analytic continuation one continues both the coordinates and the parameters of the solutions. In the continuation to Klein space, the inclusion of NUT charge in a self-dual way leads to especially simple spacetimes with interesting properties. From this point, we will mainly  restrict our discussion to the self-dual case
\begin{equation}
    M=N.
\end{equation}
The anti-self-dual case $M = - N$ is easily obtained by interchanging $t\to -t$.

\subsection{Toric Penrose Diagrams}\label{sec:22penrosediagram}

In Lorentzian signature, Penrose diagrams usefully capture the causal structure of a wide variety of spacetimes, especially those involving black holes. In Kleinian spacetimes, all points can typically be connected by everywhere spacelike, timelike, or null curves so there is no causal structure in the usual sense. However the spacetimes considered herein all have two compact Killing symmetries, one originating from  time and the other from the azimuthal angle in the original Lorentzian section. 
Hence the spacetimes are a generalized fibration of a Lorentzian torus over a Lorentzian base. Reducing along the Lorentzian torus leads to a 2D Lorentzian spacetime. The global structure of the full Kleinian spacetime can then be usefully described in terms of the ordinary Penrose diagram of the Lorentzian base together with indications of where various cycles of the Lorentzian torus 
degenerate, as is customary in (albeit typically fully Euclidean) toric geometry. This mirrors the usual Lorentzian Penrose diagram which often has a sphere at each point degenerating at various boundaries. The ``toric Penrose'' diagrams are  a hybrid of the toric diagrams of toric geometry and the Penrose diagrams of general relativity and capture some aspects of the global structure.  We begin with a review of the elementary but non-trivial example of flat Klein space described in  \cite{Atanasov:2021celesttorus}.

\subsubsection{Flat Klein Space}\label{ssec:kleinpenrose}
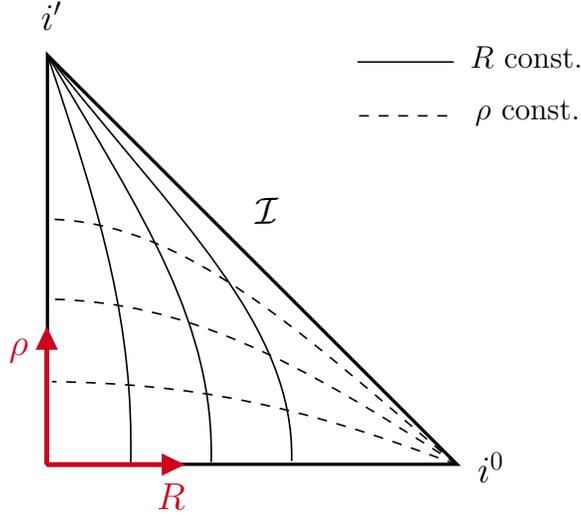
\begin{figure}
    \centering
    \resizebox{0.5\textwidth}{!}    {
        \tikzset{every picture/.style={line width=0.75pt}} 

\begin{tikzpicture}[x=0.75pt,y=0.75pt,yscale=-1,xscale=1]

\draw  [color={rgb, 255:red, 0; green, 0; blue, 0 }  ,draw opacity=1 ][line width=1.5]  (208,95.73) -- (459,346.73) -- (208,346.73) -- cycle ;
\draw    (208,95.73) .. controls (242,196.73) and (261,262.73) .. (259,346.73) ;
\draw    (208,95.73) .. controls (268.33,196.73) and (310.33,261.4) .. (308.33,345.4) ;
\draw    (208,95.73) .. controls (286.33,195.4) and (359.67,260.73) .. (357.67,344.73) ;
\draw  [dash pattern={on 4.5pt off 4.5pt}]  (458.86,347) .. controls (359.3,308.99) and (294.78,297.37) .. (210.77,296.01) ;
\draw  [dash pattern={on 4.5pt off 4.5pt}]  (459,346.73) .. controls (360.49,282.41) and (292.75,246.8) .. (208.74,245.44) ;
\draw  [dash pattern={on 4.5pt off 4.5pt}]  (458.86,347) .. controls (362.41,264.75) and (291.68,197.69) .. (207.67,196.33) ;
\draw [color={rgb, 255:red, 208; green, 2; blue, 27 }  ,draw opacity=1 ][line width=2.25]    (207.61,346.91) -- (207.61,267.2) ;
\draw [shift={(207.61,262.2)}, rotate = 450] [fill={rgb, 255:red, 208; green, 2; blue, 27 }  ,fill opacity=1 ][line width=0.08]  [draw opacity=0] (14.29,-6.86) -- (0,0) -- (14.29,6.86) -- cycle    ;
\draw [color={rgb, 255:red, 208; green, 2; blue, 27 }  ,draw opacity=1 ][line width=2.25]    (207.61,346.91) -- (287.32,346.91) ;
\draw [shift={(292.32,346.91)}, rotate = 180] [fill={rgb, 255:red, 208; green, 2; blue, 27 }  ,fill opacity=1 ][line width=0.08]  [draw opacity=0] (14.29,-6.86) -- (0,0) -- (14.29,6.86) -- cycle    ;
\draw    (398,100) -- (457,100) ;
\draw  [dash pattern={on 4.5pt off 4.5pt}]  (399,133) -- (458,133) ;

\draw (210.45,71.6) node  [font=\LARGE,color={rgb, 255:red, 0; green, 0; blue, 0 }  ,opacity=1 ]  {$i'$};
\draw (342.35,191.2) node  [font=\LARGE,rotate=-359]  {$\mathcal{I}$};
\draw (479.78,348.67) node  [font=\LARGE,color={rgb, 255:red, 0; green, 0; blue, 0 }  ,opacity=1 ]  {$i^{0}$};
\draw (190.47,276.87) node  [font=\LARGE,color={rgb, 255:red, 208; green, 2; blue, 27 }  ,opacity=1 ] [align=left] {$\displaystyle \rho $};
\draw (283.8,366.2) node  [font=\LARGE,color={rgb, 255:red, 208; green, 2; blue, 27 }  ,opacity=1 ] [align=left] {$\displaystyle R$};
\draw (501.57,97) node  [font=\Large] [align=left] {$\displaystyle R$ const.};
\draw (502.32,131) node  [font=\Large] [align=left] {$\displaystyle \rho $ const.};

\end{tikzpicture}
    }
\caption{The toric Penrose diagram for flat Klein space \cite{Atanasov:2021celesttorus}. Null lines are at 45$^\circ$, and lines of constant $\rho$ and $R$ are shown. Timelike ($i'$), spacelike ($i^0$) and null infinity ($\mathcal{I}$) are also labelled. }
\label{fig_flatPenrose}
\end{figure}

The Lorentzian Minkowski metric is
\begin{equation}
    ds_{\textrm{Mink}}^2=-dt^2+dz^2+dx^2+dy^2.
\end{equation}
There are many ways to reach flat Klein space from this via analytic continuation. Here we take 
\begin{align}
        t\to it,\,\,& x\to ix,\,\, y\to iy\,\, \label{eq:rotxy}
\end{align}
because this is the flat analog of \eqref{eq:ttoit} and \eqref{eq:thetatoitheta} and, as we will see  in section \ref{sec:amplitudes}, affords direct comparison with results from the amplitudes program. Applying \eqref{eq:rotxy} gives
\begin{align}
    ds^2_{\text{Klein}}=&dt^2+dz^2-dx^2-dy^2 \label{eq_flatmetricconf1}\\
    =& dR^2 + R^2d\psi^2 - d\rho^2 -\rho^2 d\phi^2 \label{eq_flatmetricconf}
\end{align}
where in the second line we have introduced double polar coordinates $R,\rho \geq 0$ and $\psi,\phi \in [0, 2\pi)$. The toric Penrose diagram in the $R$--$\rho$ plane is found by conformally compactifying \eqref{eq_flatmetricconf} and is shown in Fig. \ref{fig_flatPenrose}. Each point in the diagram is a Lorentzian torus generated by the two $U(1)$ cycles of $\phi$ and $\psi$. 

The conformal metric on null infinity ($\mathcal{I}$) is the product of a square  Lorentzian torus and a null interval.  The spacelike cycle of the torus degenerates along the timelike line $R=0$, while the timelike cycle degenerates along the spacelike line $\rho=0$. The restriction $R,\rho\geq 0$ is then imposed in \eqref{eq_flatmetricconf} so that the cycles are non-degenerate except at the boundary. This is the analog of the restriction \eqref{eq:thgeq0} for Taub-NUT. The identifications 
\begin{equation}
    \psi\sim \psi + 2\pi\,,\,\, \phi \sim \phi +2\pi \,
\end{equation}
prevent the metric from having conical singularities along the degenerate loci. 

\subsubsection{Self-Dual Taub-NUT}

We will now construct toric Penrose diagrams for non-rotating self-dual Kleinian black holes, which have expected black hole features like a singularity and horizon (in the reduced geometry). Other cases, namely the Kleinian Schwarzschild and $M=0$ Taub-NUT toric Penrose diagrams, are given in Appendix \ref{sec:app_schwarz}. 

The self-dual Kleinian Taub-NUT metric from \eqref{eq_22TaubNUTgeneric} takes the form
\begin{equation}
    ds_{\textrm{TN}}^2 = \frac{r-M}{r + M}(dt - 2 M \cosh \theta d \phi)^2 + \frac{r + M}{r-M}dr^2 - (r^2 - M^2)( d \theta^2 + \sinh^2 \theta d \phi^2 ), \label{eq_Klein_TN_selfdual}
\end{equation}
where $r \in [-M, \infty)$, $\theta \in [0, \infty)$ and $t$ and $\phi$ obey the periodicity conditions given in \eqref{eq_22TN_MisN_period1} and \eqref{eq_22TN_MisN_period2}. Recall also we impose $\theta\geq 0$ for non-degeneracy of the torus cycles. While the metric coefficients further degenerate at $r=\pm M$, it will be shown shortly that $r=M$ is only a coordinate singularity. Indeed, the Kretschmann scalar \eqref{eq_22TNKretsch} reduces to
\begin{equation}
 R_{\mu \nu \rho \sigma} R^{\mu \nu \rho \sigma} = \frac{96 M^2}{(r + M)^6},
\end{equation}
signaling that only the singularity at $r=-M$ survives in the self-dual case. 

\begin{figure}
    \centering
    \resizebox{0.9\textwidth}{!}    {
        \input{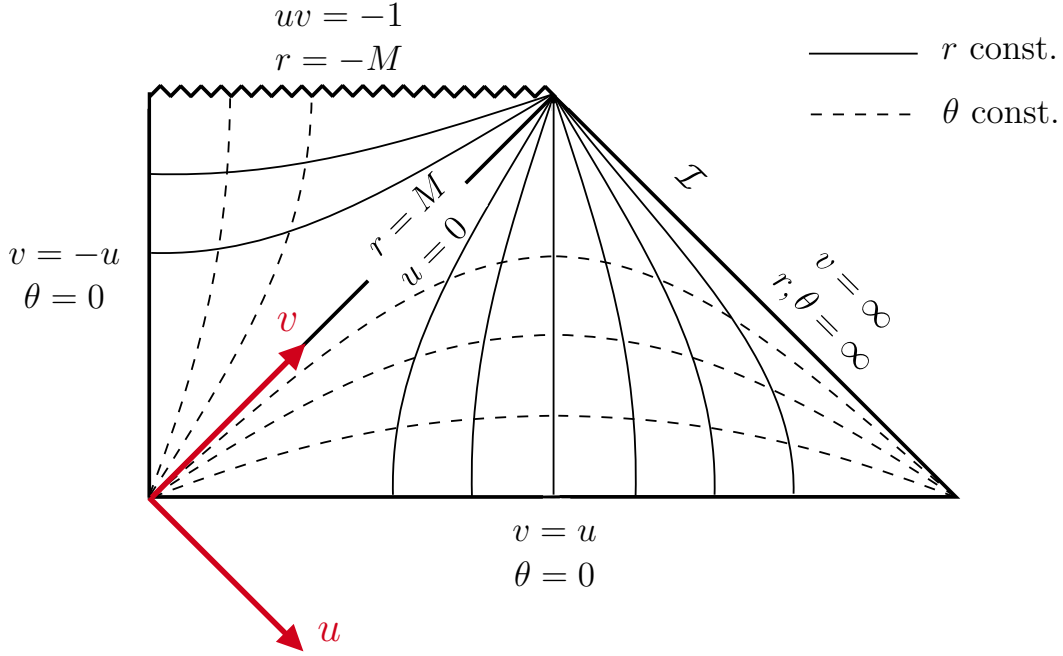}
    }
    \caption{The toric Penrose diagram for Kleinian $M=N$ Taub-NUT. There is a horizon at $u=0$, and a singularity at $uv = -1$, corresponding to $r=-M$.  Lines of constant $r$ are solid, and lines of constant $\theta$ are dashed. Null infinity, $\mathcal{I}$, and the horizon $r=M$ are labelled. The timelike cycle of the torus degenerates along the spacelike line $v=u$, while the spacelike cycle degenerates along the timelike $v=-u$ line.}
    \label{fig_SDTaubNUT}
\end{figure}

Assuming $M>0$, $t$ and $\phi$ are periodic compact coordinates. So, we can construct a toric Penrose diagram in $r$ and $\theta$, with $t$ and $\phi$ surpressed at each point. Taking $t$ and $\phi$ to be constant yields the $(1,1)$ metric
\begin{equation}
    ds_{2\textrm{d}}^2 =\frac{r+M}{r-M} dr^2  - (r^2 - M^2)d\theta^2. \label{eq_11TNSD}
\end{equation}
To construct the maximal extension of the metric, we introduce the coordinate change 
\begin{gather}
    r = M(2uv+1)\\
    \theta = \tanh^{-1}\left(\frac{ v^2 - u^2 }{ v^2 + u^2 }\right),
\end{gather}
at which point the metric \eqref{eq_11TNSD} becomes
\begin{equation}
    ds_{2\textrm{d}}^2 = 16M^2 (1 + uv) du dv. 
\end{equation}
Using the non-degeneracy condition $\theta \geq 0 $, the $u$ and $v$ coordinates satisfy $|v| \geq |u|$, which yields two disconnected spacetime copies. We choose the $v \geq |u|$ region. We now carve out the space by the singularity at $uv=-1$. The resulting toric Penrose diagram is shown in Fig. \ref{fig_SDTaubNUT}. It can furthermore be checked that null infinity $\mathcal{I}$ and the horizon $r=M$ are both geodesically complete.

We also note that constant $r$ slices of this metric are warped $AdS_3$. Indeed, the warped $AdS_3$ metric is
\begin{equation}
    ds_{{WAdS}_3}^2 = - \frac{1}{4}F( d \psi - \cosh \theta d \phi)^2+ \frac{1}{4}(d \theta^2 + \sinh^2 \theta d \phi^2 ),\label{eq_warpedAdS}
\end{equation}
where we refer to $F$ as the ``warping factor,'' since the $F = 1$ case is unwarped $AdS_3$. In the metric \eqref{eq_Klein_TN_selfdual}, upon setting $t = 2M \psi$, constant $r$ slices are warped $AdS_3$ with warping factor 
\begin{equation}
    F = \frac{4 M^2}{(r+M)^2} = \frac{1}{(1+uv)^2}. \label{eq_warping factor}
\end{equation}
Finally, we consider the $\phi$ and $t$ cycles in more detail, since they are suppressed in the toric Penrose diagram. The coordinates $\phi$ and $t$ are compact timelike and spacelike directions which generate cycles on a Lorentzian torus. Upon crossing the horizon $r=M$, $\phi$ and $t$ switch spacelike and timelike character. In Appendix \ref{app_torus}, the modular parameter of the torus is calculated to be
\begin{equation}
    \tau = \left(1-\frac{2F}{1+F +(F-1)\cosh\theta} \right) + i\left(\frac{2\sqrt{F}\tanh{\theta/2}}{1+F +(F-1)\cosh\theta}\right).\label{eq_torusmodparam}
\end{equation}

The timelike cycle of the torus degenerates along the spacelike line $v=u$, while the spacelike cycle degenerates along the timelike $v=-u$ line. In terms of the modular parameter, along $\theta = 0$ we have $\tau = 0$. On $\mathcal{I}$ we have $r, \theta = \infty$ and $\tau = 1.$

\subsubsection{Flat Limit of Kleinian  Taub-NUT }\label{sec:rectns}

One may wonder how the Taub-NUT metric with $M = N = 0$ relates to flat Klein space discussed in section \ref{ssec:kleinpenrose}. Consider the rectangular coordinate system \eqref{eq_flatmetricconf1}. It is useful to consider the patch
\begin{equation}
    z>\sqrt{x^2+y^2}\geq 0 \label{eq:wedge}
\end{equation}
which corresponds to a Rindler wedge of Klein space. (In this patch, we will be able to express Kleinian black holes in terms of scattering amplitudes in section \ref{sec:amplitudes}.) Let us introduce the parameterization
\begin{equation}
\begin{split}
    z& =\,r \cosh \theta, \\
    x+i y &= r e^{i\phi} \sinh \theta\,, \label{eq:someco}
\end{split}
\end{equation}
where $r = \sqrt{z^2 -x^2 -y^2}> 0$. (Setting $\rho = r \sinh \theta$ gives the same $\rho$ as in the flat Klein metric \eqref{eq_flatmetricconf}, where the condition $\rho\geq0$ yields $\theta\geq 0$). With this parameterization, \eqref{eq_flatmetricconf} takes the form 
\begin{equation}
    ds^2_{\text{Klein}} =dt^2 + dr^2 - r^2 ( d \theta^2 + \sinh^2 \theta d \phi^2 )\,,
\end{equation}
which is precisely the $M=N=0$ Taub-NUT metric \eqref{eq_22TaubNUTgeneric}. We can therefore see that the $(2,2)$ Taub-NUT coordinates are only continuously connected to a Rindler wedge of Klein space. We also note that the continuation $(x,y) \to (ix, iy)$ is equivalent to the continuation $\theta \to i \theta$. This is because one can start from the standard Lorentzian flat space spherical coordinates and then send $\theta \to i \theta$; from $\cos(i \theta) = \cosh \theta$ and $\sin(i \theta) = i \sinh \theta$, the resulting parameterization will Wick rotate $x$ and $y$, and yield the $x$ and $y$ parameterizations of \eqref{eq:someco}. So, we have
\begin{equation}
    (x,y)\to (i x,i y)\,\, \Longleftrightarrow \,\,\theta \to i\theta \,.\label{eq:corrrot}
\end{equation}
 
Together with $t\to it$, $\theta \to i \theta$ is the same as the Taub-NUT rotation \eqref{eq:ttoit}--\eqref{eq:thetatoitheta} done in the previous section. In fact, this analytic continuation will enable us to construct the linearized $(2,2)$ Taub-NUT solution from scattering amplitudes in section \ref{sec:amplitudes}. This linearized solution will be defined on the Rindler wedge.

However, it should be noted that the construction of the Taub-NUT toric Penrose diagram above required a finite $M\neq 0$, and thus proceeded differently than in flat space. The main difference is that the time direction is non-compact in the $M=0$ case. Indeed, the identifications \eqref{eq_22TN_MisN_period1}--\eqref{eq_22TN_MisN_period2} established for Taub-NUT cease to constrain $t$ and reduce to the expected condition $\phi \sim \phi+2\pi$ for flat space. Having said that, in the flat case a new compact direction arises when one considers the maximal extension beyond this Rindler wedge.

\subsection{Kerr-Taub-NUT  as diffeomorphisms of Taub-NUT}\label{sec:largediffeoKTN}

One might wonder what happens to Kleinian Taub-NUT spaces when an angular momentum parameter $a$ is introduced. The corresponding metrics were studied by Plebanski and Demianski as a subset of a 5-parameter family of solutions to the Einstein equations \cite{Plebanski:1975xfb,Plebanski:1976gy}. They exhibit interesting features such as integrability \cite{Chong:2004hw} and recently observed double-copy relations to gauge theory \cite{Luna:2015paa,Luna:2018dpt}.

Here we will consider the Kerr-Taub-NUT metric, which is determined by three parameters $(M,N,a)$. We will construct its Kleininan signature metric following the analytic continuation used in the previous section. 

Surprisingly, in the $M = N$ case, we will show that the parameter $a$ can be eliminated by a diffeomorphism. It should be noted that this does not preclude the Kerr-Taub-NUT metric from having angular momentum because large diffeomorphisms can change the charges of the solution, as in \cite{Banados:1992gq,Strominger:2017zoo}.

In the Lorentzian Kerr-Taub-NUT metric \cite{miller1973global} both the spin $a$ and the NUT charge $N$ are odd under time reversal. Reality of the metric then requires that the analytic continuation $t\to it$ must be supplemented by $a\to i a$ and $N \to iN$. After further rotating $\theta\to i\theta$ we obtain a $(2,2)$ metric given by
\begin{align}\label{Kktn}
    ds^2 &= \Sigma (\frac{d{r'}^2}{\Delta} - d {\theta'}^2) - \frac{\sinh^2 \theta' }{\Sigma} (a dt - \rho d \phi)^2 + \frac{\Delta}{\Sigma}(dt + A d \phi)^2 \\
    \Sigma &= {r'}^2 - (N + a \cosh \theta')^2 \\
    \Delta &= {r'}^2 - 2 M r' + N^2 - a^2\\
    A &= -a \sinh^2 \theta' - 2 N \cosh \theta' \\
    \rho &= {r'}^2 - N^2 - a^2 = \Sigma - a A.
\end{align}
As in the Taub-NUT case, this metric contains two singularities. This can be seen from the Kretschmann scalar
\begin{equation}
 R_{\mu \nu \rho \sigma} R^{\mu \nu \rho \sigma} = 24 \left(\frac{(M - N)^2}{(N - r'+a\cosh\theta')^6} + \frac{(M + N)^2}{(N + r'+a\cosh\theta')^6}\right)\,. \label{eq_22KTNKretsch}
\end{equation}

When $M=\pm N$, $a$ can be absorbed into a shift of $r$ in the above expression. This suggests that the
Kerr-Taub-NUT metric could be diffeomorphic to the Taub-NUT metric. Indeed, starting from the self-dual Taub-NUT metric \eqref{eq_Klein_TN_selfdual} and applying the coordinate transformation
\begin{align}\label{large_diffeo_1}
    \theta &= 2\, \mathrm{arctanh} \left( \tanh \left( \frac{\theta'}{2}\right) \sqrt{\frac{r' - M - a}{r' - M + a}} \right) \,,\\
    r &= r' + a \cosh \theta' \,, \label{large_diffeo_2}
\end{align}
we recover the self-dual Kerr-Taub-NUT metric \eqref{Kktn}! Moreover, we can check that the toric Penrose diagrams are related by this diffeomorphism. Note that the only singularity of self-dual Kerr-Taub-NUT, located at
\begin{equation}
    r'=-M-a\cosh\theta'
\end{equation}
is now mapped to the singularity $r=-M$ of Taub-NUT. We are left to check if the horizons match. We consider the 2d metric for Kerr-Taub-NUT with $dt=d\phi=0$,
\begin{equation}
    ds^2= \Sigma \left(\frac{d{r'}^2}{\Delta} - d {\theta'}^2\right) = \Sigma (dR^2-d\theta'^2)\,,\label{eq:horizonKTNs}
\end{equation}
where $r'=M-a\cosh R$. There is a horizon located at $R=\theta'$, i.e.
\begin{equation}
    r'=M-a\cosh\theta'\,\label{eq_ktnhorizonlocation}.
\end{equation}
This is exactly the Taub-NUT horizon $r=M$, under the diffeomorphism \eqref{large_diffeo_2}. This confirms that the singularity and horizon in the toric Penrose diagrams are preserved under the diffeomorphism.

The transformation is mathematically related to the so-called Newman-Janis (NJ) algorithm in Lorentzian signature \cite{newman1965note}. However, it has crucial differences. The NJ algorithm was first used to obtain the Kerr metric from the Schwarzschild metric, by introducing a complexification of the metric $f(r)\to f(r,\bar{r})$ followed by the replacements
\begin{align}
\begin{split}
    r'  \,\to \, & r - i a\cos\theta \,, \\
    \bar{r}'  \,\to \, & \bar{r} + i a\cos\theta. \label{eq:comp}
\end{split}
\end{align}
Importantly, this is not a true change of coordinates but rather a replacement rule, as the Kerr and Schwarzschild metrics are certainly not diffeomorphic.  On the other hand, after a naive analytic continuation of \eqref{eq:comp} to $(2,2)$ signature, the Newman-Janis shift hints at taking $r'\to r + a \cosh\theta$. As shown above, for the self-dual Kerr-Taub-NUT solution we can obtain a true diffeomorphism by shifting $r$ in such a way alongside a coordinate transformation of $\theta$.

There are alternate coordinates in which the spin-shifting coordinate transformation takes a particularly simple form. Defining $(x,y,z)$ by
\begin{align}
x' +i y'& =\sqrt{(r'-M)^{2}-a^{2}}e^{i\phi'}\sinh\theta',\,\label{eq:linco}\\
z' & =(r'-M)\cosh\theta',\,
\end{align}
for Kerr-Taub-NUT and 
\begin{align}
x +i y& =(r-M)e^{i\phi}\sinh\theta\,,\label{eq:linco_b}\\
z & =(r-M)\cosh\theta\,,
\end{align}
for Taub-NUT, the diffeomorphism \eqref{large_diffeo_1} and \eqref{large_diffeo_2} can be shown to simply amount to a shift in $z$:
\begin{equation}
(x', y', z'+a) = (x, y, z).
\end{equation}

As we will see in the next section, it is in fact these simple coordinates which naturally arise from using scattering amplitudes to study black holes. The NJ shift was understood using scattering amplitudes in \cite{Arkani-Hamed:2019ymq} by connecting the classical scattering angle between a spinning black hole and a test particle to {\it four-point} scattering amplitudes. In the next section of this paper, however, we will give a different and in some respects simpler amplitudes-inspired interpretation to the NJ shift. In particular, we will connect a {\it three-point }amplitude directly to a stationary black hole metric.

\section{Stationary Spacetimes and Scattering Amplitudes}\label{sec:amplitudes}

In order to better understand these analytically continued metrics and the emergent diffeomorphism between Taub-NUT and Kerr-Taub-NUT, we will now connect our discussion of $(2,2)$ metrics to scattering amplitudes.

We will argue that stationary black hole metrics in $(2,2)$ signature, to leading order in Newton's constant $G$, are completely specified by 3-point scattering amplitudes. The amplitudes correspond to a graviton being emitted by a heavy massive particle. In $(1,3)$ signature, only the radiative data of the metric can be connected to scattering amplitudes. Since there is no radiative data for stationary sources, Lorentzian amplitudes are a priori unrelated to stationary spacetimes such as Schwarzschild and Taub-NUT. However, after analytic continuation to $(2,2)$ signature, we will show that these metrics are completely determined by radiation modes, and hence are given by scattering amplitudes.

A connection between black hole spacetimes and the classical limit of QFT particles was first put forward by Duff \cite{PhysRevD.7.2317}. He showed that the Schwarzschild solution can be obtained from a classical source via off-shell Feynman diagrams. The source can be understood as a massive particle coupled to an off-shell graviton leg, which is effectively a three-point correlation function. However, such correlation functions in general carry a large amount of unphysical information corresponding to gauge redundancies.\footnote{The corresponding correlation function for the Kerr metric has been computed via QFT form factors in \cite{Chung:2019yfs}.} 

On the other hand, on-shell  scattering amplitudes are recently proving to be extremely powerful for the  study of classical black hole spacetimes.\footnote{Especially in the context of the two-body problem in GR, see e.g. \cite{Cheung:2018wkq,Bern:2019crd,Bern:2021dqo,Bjerrum-Bohr:2018xdl,Damour:2017zjx,DiVecchia:2020ymx,Aoude:2020onz,Mogull:2020sak,Kalin:2019rwq} for some of the recent approaches.} In previous studies, a correspondence between black holes and QFT particles emerges indirectly by considering a test body in a black hole background, see e.g. \cite{Cachazo:2017jef,Kosower:2018adc,Arkani-Hamed:2019ymq, Emond:2020lwi}. In the Born approximation, the dynamics are controlled by a four-point amplitude for graviton exchange between two massive particles (one of which being the test particle). By considering observables such as the scattering angle, one can then match the dynamics of the test particle to four-point amplitudes corresponding to certain backgrounds. Then, using analyticity the four-point amplitude can be determined from its factorization into three-point amplitudes. Using this procedure, the three-point amplitudes for Kerr and Kerr-Taub-NUT have recently been written down \cite{Guevara:2018wpp,Chung:2018kqs,Emond:2020lwi,Huang:2019cja}. The spin is added by considering spinning particles minimally coupled to a graviton, whereas the NUT charge $N$ emerges naturally by introducing a parity dependence of the coupling to the graviton.

Here, on the other hand, we will establish a \textit{direct} relation between stationary spacetimes and three-point amplitudes. We will focus on the cases of Taub-NUT and Kerr-Taub-NUT for general $M,N$, which includes the Schwarzschild spacetime. We will first explain why this correspondence is not possible a priori in Lorentzian signature, but is naturally achieved via analytic continuation in Kleinian signature. We will also find that the natural analytic continuation procedure for amplitudes is the same as the one we used to construct Kleinian metrics in section \ref{sec:sec2}. This choice of analytic continuation will be further justified from the perspective of the S-matrix.

This section will largely be conducted in $(2,2)$ signature. For convenience, we define $\eta^L_{\mu\nu}$ to be the $(1,3)$ flat metric and $\eta_{\mu\nu}$ to be the $(2,2)$ flat metric:
\begin{align}
    \eta^L_{\mu \nu} &\equiv \mathrm{diag}(-,+,+,+) \\
    \eta_{\mu \nu} &\equiv \mathrm{diag}(+,-,-,+) \label{eq_kleinflatmetric}.
\end{align}
We also define
\begin{align}
    k \cdot x_L &\equiv \eta^L_{\mu \nu} k^\mu x_L^\nu 
\end{align}
in Lorentzian signature and
\begin{align}
    k \cdot x &\equiv \eta_{\mu \nu} k^\mu x^\nu \\
    k^2 &\equiv \eta_{\mu \nu} k^\mu k^\nu 
\end{align}
in Kleinian signature.

\subsection{General Setup and Connection to Amplitudes} \label{sec:ampgeneralsetup}
In this section, we will first outline how in Lorentzian signature, one can obtain the radiative part of a metric from scattering amplitudes. However, in this work, we wish to consider stationary (non-radiative) spacetimes. It turns out that the radiative part of a stationary metric vanishes for the exact same reason that Lorentzian three-point amplitudes with one graviton vanish. Given that the vanishing of three-point amplitudes is resolved by analytic continuation to Klein space, this will motivate the idea that Kleinian stationary metrics can be written in terms of three-point amplitudes. 

We start by considering the perturbative expansion of a metric around flat space, in Lorentzian signature.\footnote{In this section we restore Newton's constant  $G$ and consider metrics with a power expansion as $G\to 0$ such that the leading order is flat space. This encompasses the metrics we have discussed so far.} We denote the term linear in $G$ by $h^L_{\mu\nu}$, and write
\begin{equation}
ds_{L}^{2}=g^L_{\mu \nu} dx^\mu_L dx^\nu_L, \hspace{1 cm} g^L_{\mu \nu} = \eta^L_{\mu \nu} + h^L_{\mu \nu}(x_L) + \mathcal{O}(G^2),\label{eq:pert}
\end{equation}
where we have used rectangular coordinates $x^{\mu}_L=(t_L,x_L,y_L,z_L)$ induced by flat space. In linearized gravity $h_{\mu\nu}$ is not gauge invariant, and can indeed be modified by linearized diffeomorphisms. Here, we will follow the standard procedure (see e.g. \cite{carroll2004spacetime, rindler2006relativity} for more details) to recast $h_{\mu\nu}$ into harmonic gauge, where the linearized Einstein equations become particularly simple. We define the trace reversed perturbation as
\begin{equation}
    \bar{h}^L_{\mu \nu} \equiv h^L_{\mu \nu} -\frac{1}{2} \eta^L_{\mu \nu} h^L.
\end{equation}
By applying a linearized diffeomorphism $\delta h^L_{\mu \nu} = 2 \partial_{(\mu} \xi_{\nu)}$, the perturbation can be cast into harmonic gauge, defined by
\begin{equation}\label{eq:dharmonic}
    \partial^\mu \bar{h}^L_{\mu \nu} = 0.
\end{equation}
The linearized Einstein equations, which couple $\bar{h}^L_{\mu \nu}$ to a conserved source $\partial_\mu \mathcal{T}_L^{\mu \nu} = 0$, take the simple form
\begin{equation}
\partial^{2}\bar{h}^L_{\mu\nu}=-16\pi G\mathcal{T}^L_{\mu\nu}(x_L)\,.  \label{eq_lorEinsteineqn}
\end{equation}
One can then solve for $\bar{h}^L_{\mu \nu}$ using the retarded propagator, finding
\begin{equation}
\bar{h}^L_{\mu\nu}=- 16\pi G\int\frac{d^{4}k}{(2\pi)^{4}}\frac{  e^{i k\cdot x_L}}{(k^0 + i\epsilon)^{2} - \vec{k}^2}\mathcal{T}^L_{\mu\nu}(k)\,.\label{eq_lorhwithretprop}
\end{equation}
In order to relate to scattering amplitudes, we must consider an on-shell quantity -- the radiative field $\bar{h}^{\text{rad}, L}_{\mu\nu}$. It is defined by taking the difference between the advanced and retarded propagators, as explained in Appendix \ref{ap:radamp}, and is given by
\begin{align}
\bar{h}^{\text{rad}, L}_{\mu\nu} &\equiv 16\pi G\int\frac{d^{4}k}{(2\pi)^{3}} \, i \, \mathrm{sign}(k^0)\delta(k_0^2-\vec{k}^2)  e^{i k\cdot x_L}   \mathcal{T}^L_{\mu\nu}(k)\,, \label{eq:hradT}
\end{align}
where the integral is over the on-shell phase space of a massless particle of momentum $k^\mu$, including positive and negative frequencies $k^0$. As outlined in Appendix \ref{ap:radamp}, in Lorentzian signature the on-shell current $\mathcal T_{\mu \nu}(k)$ is in correspondence with a classical limit of a graviton emission amplitude $\mathcal{M}_{n,L}^\pm$ through
\begin{equation}
    \mathcal{M}_{n,L}^\pm = \epsilon^\pm_{\mu \nu}(k) \mathcal{T}_L^{\mu \nu} (k) \hspace{0.5 cm} \text{at } k_0^2 -\vec{k}^2 = 0\,\label{eq_lorMtoT}\,\,\, , \,\, n>3.
\end{equation}
Here $n$ corresponds to the total number of particles in the amplitude, including one outgoing graviton (the case $n=3$ is set apart as we explain in a moment).
The polarization tensors $\epsilon^\pm_{\mu \nu}$ correspond to the two helicities of the graviton. Thus, in Lorentzian signature only the radiative components of the metric \eqref{eq:hradT} can be linked to scattering amplitudes. This usually emerges when considering scattering of two or more compact objects. For more details, we refer the interested reader to \cite{Luna:2016due,Luna:2016hge,Luna:2017dtq,Kosower:2018adc,Bautista:2019tdr,Bautista:2021wfy,Manu:2020zxl,Bautista:2021inx,Cristofoli:2021vyo,Jakobsen:2021lvp,Jakobsen:2021smu,Jakobsen:2021zvh,Bautista:2019evw} where classical radiation has been studied through this correspondence.

However, in this work we want to study stationary solutions for which there is no radiation in Lorentzian signature. Indeed, let us specialize the above discussion. If we pick the time direction
vector $u^{\mu} = (1,0,0,0)$,
time translation symmetry implies that $ k^0 = 0$, so that
\begin{equation}
\mathcal{T}^L_{\mu\nu}(k)=-2\pi\delta(k^0)T^L_{\mu\nu}(k)\,,\label{eq:tsym}
\end{equation}
for some $T^L_{\mu\nu}(k)$ and \eqref{eq_lorhwithretprop} becomes the Euclidean integral (independent of $i \epsilon$ prescription)
\begin{align}
\bar{h}^L_{\mu\nu}(x)&=16\pi G\int\frac{d^{3}\vec{k}}{(2\pi)^{3}}\frac{ e^{i k\cdot x_L}  }{\vec{k}^{2}}T^L_{\mu\nu}(0, \vec{k})\,.\label{eq:heuc}
\end{align}

The source $T_{\mu\nu}$ is known explicitly for a variety of solutions: for example, for the linearized Schwarzschild solution, or a general spinless compact object of mass $M$, it is given by 
\begin{equation}
    T_{\mu\nu}^{\text{Sch},L}=M u_{\mu}u_{\nu}.\label{eq:schwr}
\end{equation}
For generic solutions, $T_{\mu\nu}$ includes an infinite series of analytic (i.e. regular) corrections in $k_{\mu}$ and intrinsic degrees of freedom such as the spin $a$ \cite{Guevara:2018wpp,Vines:2017hyw}. While the explicit forms of these
sources will not be needed in the following, we will use the property of analyticity for stationary sources.

In Lorentzian signature, a stationary source \eqref{eq:tsym} yields only non-radiative Coulomb modes. Indeed, we can directly check that 
\begin{equation}
    \bar{h}^{\text{rad}, L}_{\mu\nu} = 0,
\end{equation}
since the conditions $k^2=0$ and $k^0 = 0$ imposed by \eqref{eq:hradT} and \eqref{eq:tsym} cannot be simultaneously satisfied (as $k^2 = 0$ has no support for real momenta $k^\mu=(0,\vec{k})$). This hints at an underlying connection to three-point scattering amplitudes, which also vanish in Lorentzian signature because the exact same kinematical constraints $k^2=0$ and $u\cdot k = k^0 =0$ cannot be satisfied.\footnote{More precisely, when the momentum is complex but on-shell ($k^2=0$), stationary sources \eqref{eq:tsym} can be interpreted as $n=3$ amplitudes, where a graviton of momentum $k$ is emitted from a massive particle of momentum $p_1^\mu=Mu^\mu$ \cite{Monteiro:2020plf,Guevara:2018wpp,Chung:2019yfs,Guevara:2020xjx}. The on-shell condition $(p_1+k)^2=p_1^2=M^2$ then reduces to the stationary condition $u\cdot k=0$, whereas analyticity in $k$ simply means that they do not have kinematic singularities. Even though these amplitudes can be constructed from general principles such as locality/unitarity \cite{Arkani-Hamed:2017jhn}, they vanish for real momenta due to kinematics.}

Since the S-matrix is defined by analytically continuing to complex momenta, this strongly suggests that to connect stationary spacetimes and amplitudes, we should analytically continue to $(2,2)$ signature. Indeed, the kinematic constraints $u\cdot k=k^2=0$ have real solutions in Klein space. We will confirm our intuition in the following. Related conclusions have been reached in \cite{Monteiro:2020plf,Guevara:2020xjx}.

\subsection{Analytic Continuation}\label{subsec:sec-cont-lin}

In this section we will construct well-defined $(2,2)$ linearized metrics and recover the same analytic continuation prescription previously discussed for the non-linear metrics. Remarkably, in $(2,2)$ we find that the linearized metric, defined in the previously discussed Rindler wedge, becomes a `free field' in the sense that it is constructed purely by on-shell radiative modes.

To define the linearized metric for a stationary source in $(2,2)$ signature, we will begin with its Lorentzian analogue, \eqref{eq:heuc}. It explicitly reads
\begin{equation}
\bar{h}^L_{\mu\nu}(x_L)=16\pi G\int_{\mathbb{R}^{3}}\frac{dk_{1}dk_{2}dk_{3}}{(2\pi)^{3}}\frac{e^{i(k_{1}x_L+k_{2}y_L+k_{3}z_L)}}{k_{1}^{2}+k_{2}^{2}+k_{3}^{2}}T^L_{\mu\nu}(k)\,.\label{eq:expinte}
\end{equation}
We want to regard this integral as defined on $(t_L,x_L,y_L,z_L)\in \mathbb{C}^4$ and study its domain of convergence. In particular, as the integral is independent of $t_L$, it is invariant under the Wick rotation $t_L\to it$. Then, Klein space can be reached by rotating either two or a single space coordinate, i.e. making one of the choices
\begin{align}
   \textrm{I:}\,\,\,& (t_L,x_L,y_L,z_L)=  ( it, ix, iy,z)\,\,,\,\,
   \, \text{ or }  \\
    \textrm{II:}\,\,\,& (t_L,x_L,y_L,z_L)=  (t,x,y, iz)\,\,,\,\label{eq:options}
\end{align}
for Klein space coordinates $t,x,y,z \in\mathbb{R}$. 

We will later show that after applying option I to \eqref{eq:expinte}, the integral converges nicely in the Rindler wedge $z>\sqrt{x^2+y^2}$. Option I will be the prescription adopted throughout this paper, which via the coordinate change \eqref{eq:someco} induces the corresponding $\theta \to i \theta$ rotation for the black hole solutions. The further continuation of the charges $N,a,\ldots$ will also be reflected in terms of scattering amplitudes. On the other hand, option II would end up leading to an integral with an imaginary exponent, which would require an $i\epsilon$ prescription for convergence.

Let us then consider option I. Keeping in mind the perturbative expansion \eqref{eq:pert} we impose
\begin{equation}
    g^L_{\mu\nu}dx^\mu_L dx^\nu_L=  g_{\mu\nu}dx^\mu dx^\nu
\end{equation}
which leads to the Klein metric for flat space \eqref{eq_kleinflatmetric} as well as
\begin{equation}
    h^L_{\mu\nu}(x_L)dx^\mu_L dx^\nu_L=  h_{\mu\nu}(x) dx^\mu dx^\nu\,\,, \,   \mathcal{T}^L_{\mu\nu}(x_L)dx^\mu_L dx^\nu_L=  \mathcal{T}_{\mu\nu}(x) dx^\mu dx^\nu.  \label{eq:LTOK}
\end{equation}
These conditions ensure that the Lorentzian expression \eqref{eq:expinte} is preserved.\footnote{Since $\mathcal{M}^\pm_3=\epsilon^{\pm}_{\mu\nu}\mathcal{T}^{\mu\nu}$, one must also rescale the components of the polarizations $\epsilon^\pm_{\mu\nu}$. Indeed, after this scaling the tensors become real (as opposed to complex in Lorentzian signature). They are constructed below in equation \eqref{eq:pol}.} Then, applying the analytic continuation, the Kleinian trace-reversed linearized metric is
\begin{equation}
\bar{h}_{\mu\nu}(x)=16\pi G\int_{\mathbb{R}^{3}}\frac{dk_{1}dk_{2}dk_{3}}{(2\pi)^{3}}\frac{e^{-(k_{1}x+k_{2}y-ik_{3}z)}}{k_{1}^{2}+k_{2}^{2}+k_{3}^{2}}T_{\mu\nu}(k)\,. \label{eq_hbarklein}
\end{equation}
Now, we will perform the $k_{3}$ integral using contour integration. For stationary solutions $T_{\mu\nu}(k)$ is analytic and so has no singularities in $k^\mu$. Thus, the only poles in the integrand occur at $k_{3}=\pm i\sqrt{k_{1}^{2}+k_{2}^{2}}$. Importantly, we impose the coordinate patch\footnote{The requirement $z > 0$ allows us to close the contour as long as we can expand $T_{\mu\nu}$ in small spin. The finite spin case addressed in the next section will instead require $z\pm a>0$.}
\begin{equation}
    z>0\,,
\end{equation} 
so that we can close the contour from above and pick up the upper pole, resulting in
\begin{equation}
\bar{h}_{\mu\nu}(x)=8\pi G\int_{\mathbb{R}^{2}}\frac{dk_{1}dk_{2}}{(2\pi)^{2}}\frac{1}{\sqrt{k_{1}^{2}+k_{2}^{2}}}e^{-(k_{1}x+k_{2}y+\sqrt{k_{1}^{2}+k_{2}^{2}}z)}T_{\mu\nu}(k)\,.\label{eq:anconh}
\end{equation}
The exponent in the integrand is now real, and we will now show that it simply reduces to a Gaussian integral. To see this, it is convenient to parameterize the integration region with a new coordinate system $(\lambda_1, \lambda_2)$, related to $k_1$ and $k_2$ by
        \begin{equation}
        k_2 + ik_1=\frac{1}{2}(\lambda_2 - i \lambda_1)^2. \label{eq:changez2}
    \end{equation}
These coordinates cover the $(k_1,k_2)$ plane twice, but we can extend the integration over the full $(\lambda_1,\lambda_2)$ plane by including an extra factor of $\tfrac{1}{2}$. The result is
\begin{align}
\bar{h}_{\mu\nu}(x)=& \frac{2G}{\pi }\int d^2\lambda \,\,   e^{- k(\lambda)\cdot x}\,T_{\mu\nu}(k(\lambda))\,,\label{eq:htoT}
\end{align}
where
\begin{equation}
k^\mu(\lambda)=\left(0,\lambda_1 \lambda_2,\frac{\lambda_1^2-\lambda_2^2}{2},\frac{\lambda_1^2+\lambda_2^2}{2}\right),\label{eq:keta}
\end{equation}
which is now now an elementary Gaussian integral. Indeed, it converges when the exponent is negative-definite, which corresponds to the Rindler wedge in Klein space discussed in section \ref{sec:rectns}.

Let us pause to interpret the result \eqref{eq:htoT}. The key ingredient is the vector $k^\mu$; it is null, $k^2=0$, with respect to the Klein metric. (Indeed, this momentum is precisely a solution of the kinematic constraints $u\cdot k=k^{2}=0$ of the previous section.) Remarkably, this implies that $\bar{h}_{\mu\nu}(x)$ in \eqref{eq:htoT} is a solution of the free-field equation
\begin{equation}
  \partial^2 \bar{h}_{\mu \nu} = 0\,,
\end{equation}
in the Rindler wedge of Klein space. This is so because the current $T_{\mu\nu}(k)$ only contributes via on-shell modes. As previously seen, in $(1,3)$ signature on-shell modes exclusively correspond to radiative data and are absent for stationary configurations. On the other hand, in $(2,2)$ all the modes are source-free/radiative in the on-shell sense. Indeed, using \eqref{eq:htoT} we will be able to derive Kleinian $1/r$ potentials which in $(1,3)$ signature would be associated to non-radiative modes.

\begin{figure}[h]%
    \centering
    \subfloat[\centering ]{{ \resizebox{0.35\textwidth}{!}    {\input{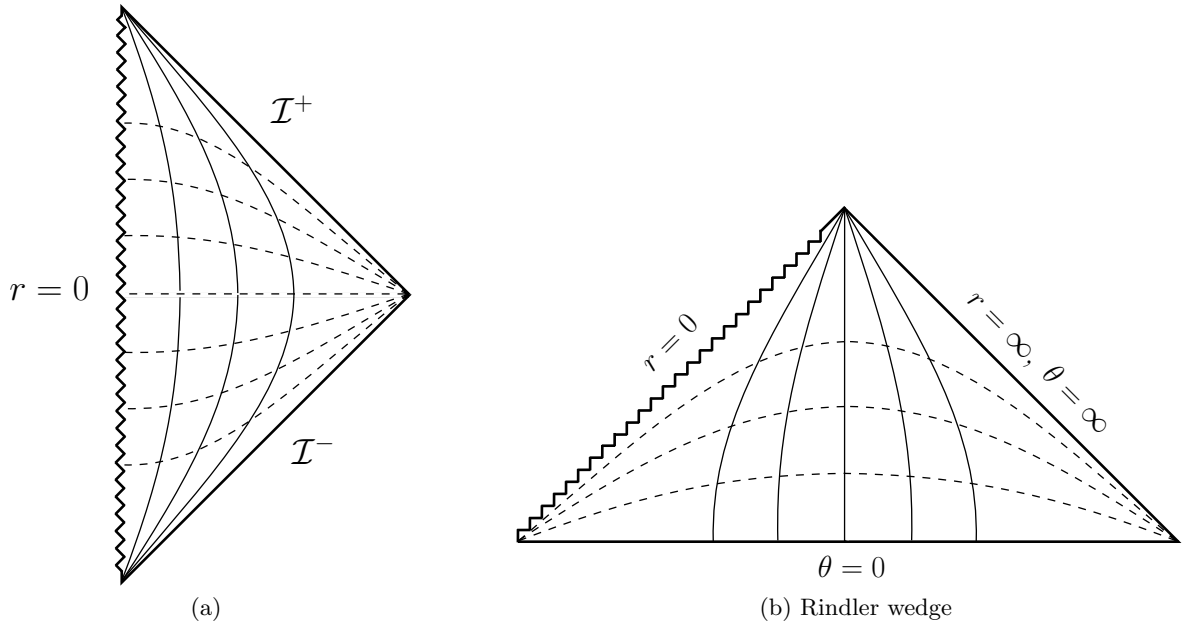} } }
    \label{fig_wedge}}%
    \qquad
    \subfloat[\centering Rindler wedge]{{ \resizebox{0.55\textwidth}{!}    {\tikzset{every picture/.style={line width=0.75pt}} 

\begin{tikzpicture}[x=0.75pt,y=0.75pt,yscale=-1,xscale=1]

\draw  [color={rgb, 255:red, 0; green, 0; blue, 0 }  ,draw opacity=1 ][line width=1.5]  (296.17,16.7) -- (547.17,267.7) -- (296.17,267.7) -- cycle ;
\draw  [color={rgb, 255:red, 255; green, 255; blue, 255 }  ,draw opacity=1 ][fill={rgb, 255:red, 255; green, 255; blue, 255 }  ,fill opacity=1 ] (294.66,20.48) -- (297.8,20.48) -- (297.8,266.18) -- (294.66,266.18) -- cycle ;
\draw [color={rgb, 255:red, 255; green, 255; blue, 255 }  ,draw opacity=1 ][line width=3]    (295.77,26.81) -- (295.87,265.59) ;
\draw [line width=1.5]    (222.78,267.24) -- (228.67,267.2) ;
\draw  [color={rgb, 255:red, 255; green, 255; blue, 255 }  ,draw opacity=1 ][fill={rgb, 255:red, 255; green, 255; blue, 255 }  ,fill opacity=1 ] (220.5,257.8) -- (234,271.3) -- (222,283.3) -- (208.5,269.8) -- cycle ;
\draw [line width=1.5]    (258.9,59.43) -- (258.94,49.48) ;
\draw [line width=1.5]    (259.25,59.08) -- (249.3,59.12) ;
\draw [line width=1.5]    (249.85,68.48) -- (249.9,58.53) ;
\draw [line width=1.5]    (250.21,68.12) -- (240.25,68.17) ;
\draw [line width=1.5]    (240.66,77.67) -- (240.71,67.72) ;
\draw [line width=1.5]    (241.02,77.31) -- (231.07,77.36) ;
\draw [line width=1.5]    (231.62,86.72) -- (231.66,76.76) ;
\draw [line width=1.5]    (231.97,86.36) -- (222.02,86.4) ;
\draw [line width=1.5]    (277.09,41.25) -- (277.09,32) ;
\draw [line width=1.5]    (277.44,40.89) -- (267.49,40.93) ;
\draw [line width=1.5]    (268.04,50.29) -- (268.08,40.34) ;
\draw [line width=1.5]    (268.4,49.94) -- (258.44,49.98) ;
\draw [line width=1.5]    (204.19,114.42) -- (204.23,104.47) ;
\draw [line width=1.5]    (204.55,114.07) -- (194.59,114.11) ;
\draw [line width=1.5]    (195.14,123.47) -- (195.19,113.52) ;
\draw [line width=1.5]    (195.5,123.11) -- (185.55,123.16) ;
\draw [line width=1.5]    (185.96,132.66) -- (186,122.71) ;
\draw [line width=1.5]    (186.31,132.3) -- (176.36,132.35) ;
\draw [line width=1.5]    (222.52,96.09) -- (222.57,86.14) ;
\draw [line width=1.5]    (222.88,95.74) -- (212.92,95.78) ;
\draw [line width=1.5]    (213.33,105.28) -- (213.38,95.33) ;
\draw [line width=1.5]    (213.69,104.93) -- (203.73,104.97) ;
\draw [line width=1.5]    (139.83,178.04) -- (139.87,168.08) ;
\draw [line width=1.5]    (140.18,177.68) -- (130.23,177.72) ;
\draw [line width=1.5]    (130.78,187.08) -- (130.82,177.13) ;
\draw [line width=1.5]    (131.14,186.73) -- (121.18,186.77) ;
\draw [line width=1.5]    (176.53,141.61) -- (176.58,131.66) ;
\draw [line width=1.5]    (176.89,141.26) -- (166.94,141.3) ;
\draw [line width=1.5]    (167.2,150.66) -- (167.25,140.71) ;
\draw [line width=1.5]    (167.56,150.3) -- (157.61,150.35) ;
\draw [line width=1.5]    (158.01,159.85) -- (158.06,149.89) ;
\draw [line width=1.5]    (158.37,159.49) -- (148.42,159.54) ;
\draw [line width=1.5]    (148.97,168.9) -- (149.01,158.94) ;
\draw [line width=1.5]    (149.32,168.54) -- (139.37,168.58) ;
\draw [line width=1.5]    (84.9,231.93) -- (84.94,221.98) ;
\draw [line width=1.5]    (85.25,231.58) -- (75.3,231.62) ;
\draw [line width=1.5]    (75.85,240.98) -- (75.9,231.03) ;
\draw [line width=1.5]    (76.21,240.62) -- (66.25,240.67) ;
\draw [line width=1.5]    (66.66,250.17) -- (66.71,240.22) ;
\draw [line width=1.5]    (67.02,249.81) -- (57.07,249.86) ;
\draw [line width=1.5]    (57.62,259.22) -- (57.66,249.26) ;
\draw [line width=1.5]    (57.97,258.86) -- (48.02,258.9) ;
\draw [line width=1.5]    (121.61,195.51) -- (121.65,185.56) ;
\draw [line width=1.5]    (121.96,195.15) -- (112.01,195.2) ;
\draw [line width=1.5]    (112.27,204.56) -- (112.32,194.6) ;
\draw [line width=1.5]    (112.63,204.2) -- (102.68,204.24) ;
\draw [line width=1.5]    (103.09,213.75) -- (103.13,203.79) ;
\draw [line width=1.5]    (103.44,213.39) -- (93.49,213.43) ;
\draw [line width=1.5]    (94.04,222.79) -- (94.08,212.84) ;
\draw [line width=1.5]    (94.4,222.44) -- (84.44,222.48) ;
\draw [line width=1.5]    (48.59,268.2) -- (48.64,258.25) ;
\draw [line width=1.5]    (48.59,267.7) -- (296.17,267.7) ;
\draw  [color={rgb, 255:red, 255; green, 255; blue, 255 }  ,draw opacity=1 ][fill={rgb, 255:red, 255; green, 255; blue, 255 }  ,fill opacity=1 ] (242.68,14.4) -- (251.08,14.4) -- (251.08,31.6) -- (242.68,31.6) -- cycle ;
\draw [line width=1.5]    (276.67,33.7) -- (295.67,14.7) ;
\draw  [color={rgb, 255:red, 255; green, 255; blue, 255 }  ,draw opacity=1 ][fill={rgb, 255:red, 255; green, 255; blue, 255 }  ,fill opacity=1 ] (295.43,18) -- (315.3,38) -- (295.43,38) -- cycle ;
\draw    (295.43,16) .. controls (329.05,116.87) and (347.83,182.79) .. (345.86,266.68) ;
\draw    (295.27,16.05) .. controls (354.92,116.92) and (396.61,183.46) .. (394.63,267.35) ;
\draw    (295.27,16.05) .. controls (261.65,116.92) and (242.87,182.84) .. (244.84,266.73) ;
\draw    (295.27,16.05) .. controls (235.62,116.92) and (194.09,183.51) .. (196.07,267.4) ;
\draw  [dash pattern={on 4.5pt off 4.5pt}]  (545.47,266.95) .. controls (447.03,228.99) and (377.59,217.43) .. (294.53,216.08) ;
\draw  [dash pattern={on 4.5pt off 4.5pt}]  (49.23,267) .. controls (147.67,229.04) and (211.46,217.43) .. (294.53,216.08) ;
\draw  [dash pattern={on 4.5pt off 4.5pt}]  (545.6,266.68) .. controls (448.2,202.44) and (381.23,166.87) .. (298.16,165.52) ;
\draw  [dash pattern={on 4.5pt off 4.5pt}]  (545.47,266.95) .. controls (450.1,184.8) and (380.17,117.83) .. (297.1,116.47) ;
\draw  [dash pattern={on 4.5pt off 4.5pt}]  (49.1,266.73) .. controls (146.5,202.49) and (213.47,166.92) .. (296.54,165.56) ;
\draw  [dash pattern={on 4.5pt off 4.5pt}]  (49.23,267) .. controls (144.6,184.85) and (214.53,117.88) .. (297.6,116.52) ;
\draw    (295.27,16.05) -- (295.22,267.29) ;

\draw (300.56,286.65) node  [font=\LARGE]  {$\theta =0$};
\draw (162.6,116.75) node  [font=\LARGE,rotate=-315]  {$r=0$};
\draw (441.7,132.85) node  [font=\LARGE,rotate=-45]  {$r=\infty ,\ \theta =\infty $};

\end{tikzpicture} } }
    \label{fig_wedge1}  }%
    \caption{(\ref{fig_wedge}) Shows Lorentzian Minkowski space with a source located at $x_L{=}y_L{=}z_L{=}0$ (corresponding to $r=0$). Lines of constant $r$ are solid, and lines of constant time are dashed.\\
    (\ref{fig_wedge1}) Shows the Rindler wedge corresponding to the linearized (2,2) Taub-NUT solution. Lines of constant $r$ are solid, and lines of constant $\theta$ are dashed. The singularity is at $r=\sqrt{z^2{-}x^2{-}y^2}=0$.}
    \label{fig_wedgeandschwarz}%
\end{figure}

An explanation why the field is source-free is as follows. In usual Minkowski space, the Lorentzian integral \eqref{eq:expinte} has a UV singularity at the origin $x_L{=}y_L{=}z_L{=}0$. This corresponds to a source at $r=0$ that propagates to all points in the space, see Fig. \ref{fig_wedgeandschwarz}a. On the other hand, \eqref{eq:htoT} was obtained via analytic continuation to Klein space by requiring, by construction, a convergent integral. This means that there is no singularity or source in the region where the integral is defined. As previously mentioned, this region is exactly the Rindler wedge where the Kleinian Taub-NUT spacetime can be perturbatively constructed, see Fig. \ref{fig_wedgeandschwarz}b.

Note that we can still ask about the singularities of \eqref{eq:htoT} -- for Taub-NUT and Schwarzschild we will show they lie at the boundary of the Rindler wedge. This corresponds to $r{=}\sqrt{z^2{-}x^2{-}y^2}{=}0$, but in contrast to Minkowski space, it is not associated to a source at $x{=}y{=}z{=}0$. This is consistent with the findings of \cite{Monteiro:2020plf} and provides a reason why the gravitational field is effectively free in the wedge. Next, we will perform a free mode expansion of the gravitational field. Later on, we will identify each mode with a scattering amplitude. 

Let us now derive such a mode expansion. Since the modes of $T_{\mu\nu}(k)$ are on-shell, we can cast it into the transverse-traceless (TT) gauge. In TT gauge, $T_{\mu\nu}(k)$ is particularly simple, as its only non-trivial modes will correspond to the two helicities of the free graviton. The gauge is imposed following a standard procedure: first, we introduce two polarization vectors such that
\begin{equation}
\epsilon^\pm(k)\cdot \epsilon^\pm(k)=0\,,\,\,\,\, \epsilon^{+}(k)\cdot\epsilon^{-}(k)=1\,,\quad\epsilon^{\pm}(k)\cdot k=0\,\,\,\,.\label{eq:pol}
\end{equation}
These vectors are real in $(2,2)$ signature. The redundancy $\epsilon^{+}\to \alpha \epsilon^{+}$, $\epsilon^{-}\to\epsilon^{-}/\alpha $
for $\alpha \in\mathbb{R}$ can be identified as a little-group $SO(1,1)$
transformation associated to the null vector $k^{\mu}$. In a Lorentz frame where $\lambda_1=\lambda_2$ in \eqref{eq:keta} we can take $\epsilon^{\pm}=\frac{1}{\sqrt{2}}(1,0,\pm 1,0)$ (so that $\epsilon^{\pm}(k)\cdot k=0$ is satisfied) and then boost it for general $\lambda_1,\lambda_2$. As the boost is orthogonal to the time direction $u^\mu$, we can always take
\begin{equation}
    \epsilon^\pm \cdot u = \frac{1}{\sqrt{2}}\,,\label{eq:epsud}
\end{equation}
for all $k(\lambda)$. This fixes the little group.

To finish imposing TT gauge, we note that the source is already transverse, since the conservation condition $\partial_\mu T^{\mu \nu} = 0$ becomes $k^{\mu}T_{\mu\nu}=0$ in momentum space. Then a
standard argument\footnote{More precisely, the condition $k^{\mu}T_{\mu\nu}=0$ leaves 6 independent
components in the source. We then consider the transformation $\delta T_{\mu\nu}(k)=2k_{(\mu}\xi_{\nu)}-\eta_{\mu\nu}k\cdot\xi$
which induces a linearized diffeomorphism in $h_{\mu\nu}$ through
equation (\ref{eq:htoT}), while still preserving the harmonic gauge (\ref{eq:dharmonic})
if $k^{2}=0$. This leaves only $6-4=2$ components in the redefined
source which we can take as $\epsilon_{\mu}^{\pm}\epsilon_{\nu}^{\pm}T^{\mu\nu}(k)$.} shows that $T_{\mu\nu}$ carries only two components, i.e. we can pick a gauge such that

\begin{equation}
T_{\mu\nu}(k)= i\epsilon_{\mu}^{+}\epsilon_{\nu}^{+}a_-(k)+i\epsilon_{\mu}^{-}\epsilon_{\nu}^{-}a_+(k)\,,\label{eq:trep}
\end{equation}
where the two modes
\begin{equation}
   i a_{\pm}(k)\equiv\epsilon_{\mu}^{\pm}\epsilon_{\nu}^{\pm}T^{\mu\nu}(k)\,\,,k^{2}=0\,,
\end{equation}
are the two polarizations associated to the $k^\mu$ direction given in $\eqref{eq:keta}$. Since the polarizations \eqref{eq:pol} are null, the current \eqref{eq:trep} is traceless. Then, in this gauge $\bar{h}_{\mu\nu}$ is a traceless field (i.e. $\bar{h}=0$) so we can obtain the full linearized metric $h_{\mu\nu}$ as 
\begin{equation}
h_{\mu\nu}(x)=\bar{h}_{\mu\nu}-\frac{1}{2}\eta_{\mu\nu}\bar{h}=\frac{2G}{\pi}\int d^2 \lambda e^{-k\cdot x}i[\epsilon_{\mu}^{+}\epsilon_{\nu}^{+}a_-(k)+\epsilon_{\mu}^{-}\epsilon_{\nu}^{-}a_+(k)]\,,\label{eq:hfroma}
\end{equation}
This is the main result of this section. It is tempting to interpret this formula as excitations of Klein space propagating freely in the outgoing $k^\mu$ direction. In $(1,3)$ spacetimes perturbations of the vaccum admit two consistent mode expansions, corresponding to asymptotic positive and negative frequency modes at $\mathcal{I}^+$ and $\mathcal{I}^-$. In contrast, in $(2,2)$ spacetimes the existence of an S-vector associated to a single component of null infinity  $\mathcal{I}$ suggests there should be a single unique mode expansion. Indeed, we have found such a mode expansion, which holds classically in the Rindler wedge of Klein space. The modes can be interpreted as having imaginary frequencies. By continuing to option II in \eqref{eq:options}, via $x^\mu \to i x^\mu$, one can obtain real frequencies, but the integral is oscillatory and requires an $i\epsilon$ prescription. 
For non-stationary solutions, mode expansions such as \eqref{eq:hfroma} will in principle have extra contributions from non-trivial singularities of $T_{\mu\nu}(k)$. We leave this for future investigations.\footnote{For more than one massive body, the source contains additional poles at $u_i\cdot k=0$. This is evident from the Braginsky-Thorne result \cite{braginsky1987gravitational}, or equivalently, from the soft factor of $n>3$ amplitudes.}

Next, we will explicitly evaluate the linearized Kleinian metric \eqref{eq:hfroma} using the classical limit of scattering amplitudes.

\subsection{Linearized Metrics and Klein Kerr-Taub-NUT from Amplitudes}\label{sec:linfromamp}

In this section, we will show that in $(2,2)$ signature stationary metrics can be obtained from a 3-point amplitude $\mathcal{M}_3^\pm$. We will explicitly check this for the $\mathcal{O}(G)$ part of the Kleinian Kerr-Taub-NUT metric of section \ref{sec:sec2}. We will also find that the existence of the diffeomorphism between self-dual Kerr-Taub-NUT and self-dual Taub-NUT is immediately apparent at the level of scattering amplitudes, as the former results from the latter by a simple shift $z \to z - a$. 

The Lorentzian relation between scattering amplitudes and on-shell classical sources, given in \eqref{eq_lorMtoT}, is trivially true for $n=3$ since both sides of the equation vanish. However, for analytically continued momenta one can argue that the (non-trivial) $n=3$ relation holds. That is, 

\begin{equation}
\mathcal{M}_{3}^{\pm}(k)\equiv\epsilon_{\mu}^{\pm}\epsilon_{\nu}^{\pm}T^{\mu\nu}(k)=ia_\pm(k) \,\,,\text{with }k^{2}=0\,,\label{eq:cir}
\end{equation}
where have used \eqref{eq:trep} to obtain the last equation and we note that both the amplitudes $\mathcal{M}_{3}^{\pm}(k)$ and the modes $a_\pm(k)$ have support on real momenta in $(2,2)$ signature. In \cite{Monteiro:2020plf}, $\mathcal{M}_{3}^{\pm}(k)$ were shown to correspond to the classical limit of graviton emission scattering amplitudes, where $\pm$ indicates the helicity. 

The correspondence \eqref{eq:cir} can be explicitly shown for several interesting cases. For instance, the two modes of the Schwarzschild source \eqref{eq:schwr}, for some real null momenta satisfying $k\cdot u=0$ and polarizations \eqref{eq:pol}, are

\begin{equation}
  \mathcal{M}_{3}^{\pm}= \epsilon^{\mu\nu}_\pm (k) T_{\mu\nu}^{\text{Sch}}=M (\epsilon^{\pm}\cdot u)^{2}\,. 
\end{equation}
These can be obtained by considering the amplitude of a scalar
massive particle $\phi$ with momentum $p^{\mu}=Mu^{\mu}$, coupled to a graviton through $h^{\mu\nu}\mathcal{T}_{\mu\nu}=h^{\mu\nu} \partial_\mu\phi \partial_{\nu}\phi  $.

Moreover, \eqref{eq:cir} has further been checked explicitly for solutions with spin $a$ in \cite{Guevara:2018wpp}.\footnote{There, both the three-point amplitudes and the classical source were supported on complex kinematics.} In this case, the spin is parameterized by the classical limit of the Pauli-Lubanski vector $a^\mu$, aligned with the $z$-axis as
\begin{equation}
    a^\mu= (0,0,0,a).
\end{equation}
Finally, we can also endow these amplitudes with NUT charge, as was shown in \cite{Huang:2019cja,Emond:2020lwi}. For the rotation $a \to i a, N\to iN$, such amplitudes are given by 
\begin{equation}
\mathcal{M}_{3}^{\pm}=(M\pm N)e^{\mp k\cdot a}(\epsilon^{\pm}\cdot u)^{2}\label{eq:sdm3}
\end{equation}
Indeed, one can see from this expression that both $a$ and $N$ are odd under time reversal, just as in the case of the spacetimes of section \ref{sec:sec2}. This is because time reversal $t\to -t$ not only induces $N\to -N, a\to -a$ but also flips the helicity of the graviton. Since self-duality of the metric is correlated with the graviton helicity, $\mathcal{M}_{3}^{+}$ and $\mathcal{M}_{3}^{-}$ will lead to self-dual and anti-self-dual metrics, respectively.

Now, the main result \eqref{eq:hfroma} of section \ref{subsec:sec-cont-lin} can be recast as a correspondence between vacuum spacetimes and scattering amplitudes. Inserting \eqref{eq:cir} into \eqref{eq:hfroma}, we find
\begin{equation}
h_{\mu\nu}^{\pm}(x)=\frac{2G}{\pi}\int d^2\lambda e^{-k\cdot x}\epsilon_{\mu}^{\mp}\epsilon_{\nu}^{\mp}\mathcal{M}_{3}^{\pm}(k)\,,\label{eq:hfromM}
\end{equation}
where $h_{\mu\nu}=h_{\mu\nu}^{+}+h_{\mu\nu}^-$. This splits the metric into self-dual ($h_{\mu\nu}^{+}$) and anti-self-dual ($h_{\mu\nu}^{-}$) pieces, which are each completely specified by a three-point amplitude. 

Equipped with the relation \eqref{eq:hfromM} between stationary spacetimes and scattering amplitudes, we will now explicitly check that the linearized (Kerr) Taub-NUT metric of section \ref{sec:sec2} can be obtained from the amplitudes \eqref{eq:sdm3}.\footnote{It's possible that, with a judicious coordinate choice, the linearized solution in fact becomes the full nonlinear solution in the self-dual case, but we have not shown this.} However, in deriving \eqref{eq:hfromM} we made particular gauge choices which change the coordinates of the linearized metric, so that the coordinates no longer match with those of the non-linear solution. There are two ways to bypass this coordinate ambiguity. In the most direct way, we can simply compute and compare curvature tensors (or their scalar components). Indeed, in Appendix \ref{app_spinor}, we will derive a closed form for the curvature of \eqref{eq:hfromM} using spinor-helicity variables. On the other hand, we can also check that \eqref{eq:hfromM} matches with (Kerr) Taub-NUT using a very simple trick as follows. Consider the gravitational potential
\begin{equation}
    g_{00} = 1- \phi^+ - \phi^- + \mathcal{O}(G^2)\,\,
\end{equation}
where
\begin{equation}
 \phi^{\pm}(x)=u^\mu u^\nu h^{\pm}_{\mu\nu}\,.
\end{equation}
For stationary spacetimes, $g_{00}$ is invariant up to a scale since it corresponds to the norm of the $\partial_t$ Killing vector. Since this scale is fixed by the flat limit, we see that the gravitational potential is a scalar under linearized gauge transformations. Moreover, one can explicitly check that the gauge transformations used to obtain \eqref{eq:hfromM} are time-independent, and hence also leave $h_{00}$ invariant! Thus, to match our metric to Kerr-Taub-NUT at the linearized level, we can simply match the values of $\phi^{\pm}$.\footnote{Note that for the self-dual pieces of the Weyl tensor the only independent components are $ C^{\pm}_{0i0j}=  \frac{1}{2}\partial_i \partial_j h^{\pm}_{00}$. 
This means that in stationary spacetimes the linearized curvature is completely fixed by the gravitational potential. }

Now, by inserting the amplitudes \eqref{eq:sdm3} into the expression for $h_{\mu\nu}$ \eqref{eq:hfromM}, and further using \eqref{eq:epsud}, the gravitational potential is given by

\begin{equation}
    \phi^{\pm} (x)= \frac{G (M\pm N)}{2\pi}\int d^2\lambda e^{-k\cdot (x\pm a)} \label{eq:phipot}\,. 
\end{equation}
This is simply a 2d Gaussian integral. To evaluate this integral, we can use the parameterization \eqref{eq:keta} to write
\begin{equation}
    k\cdot (x\pm a) =\frac{1}{2} \lambda^A \lambda^B Q_{AB}^\pm \,\,\,,\,\,A,B=1,2\,,
\end{equation}
where $Q_{AB}^\pm$ is a quadratic form given by

\begin{equation}
       Q^\pm_{AB}\equiv \left(\begin{array}{cc}
 z \pm a +y & x\\
  x & z \pm a-y 
\end{array}\right)\,. \label{eq:bigX}
\end{equation}
The two eigenvalues will be positive and the integrals will converge if and only if
\begin{align}
    \det(Q_{AB}^\pm)&=(z \pm a)^2-x^2-y^2 > 0 \,,\\
     \tr(Q_{AB}^\pm)&=z \pm a > 0\,.
\end{align}
For small spin $a\to 0$, the second condition is in fact the one that was required to close the contour in \eqref{eq:anconh}. Furthermore, in such cases the convergence of the integral \eqref{eq:phipot} translates to
\begin{equation}
    z>0\,\,\, \textrm{and}\,\,\, z^2 >x^2+ y^2
\end{equation}
So, the Taub-NUT solution can indeed be linearized in the Rindler wedge, as discussed in section \ref{sec:rectns}. We note that in general, the singularity of \eqref{eq:phipot} corresponds to the vanishing of the determinant of $Q^\pm_{AB}$, so the metric cannot be defined via integration there. Returning now to the integral \eqref{eq:phipot}, the result is given by
\begin{align}
   \phi^{\pm}(x)= \frac{G(M\pm N)}{\sqrt{(z \pm a)^2 - x^2 - y^2}}. \label{eq:pmphi}
\end{align}
In the Taub-NUT case $a=0$, one recovers the Newtonian potential $\phi^++\phi^-=2GM/r$ where $r=\sqrt{z^2 -x^2-y^2}$. This result confirms our previous claim that even Coulomb modes can be captured by on-shell amplitudes in $(2,2)$ signature. 

Now, all that is left to do is to compare this result with the gravitational potentials obtained from the full, non-linearized Kerr
Taub-NUT metric given in \eqref{Kktn}. Restoring factors of $G$, we find the non-linear result
\begin{align}
g_{00} & =
\frac{\Delta-a^2\sinh^{2}\theta}{\Sigma}\nonumber \\
 & =\frac{r^{2}-2GMr+G^{2}N^{2}-a^{2}\cosh^{2}\theta}{r^{2}-(GN+a\cosh\theta)^{2}}\,,
\end{align}
It is easy to extract the self-dual and anti-self-dual components if we cast this expression into a more suggestive form (with components proportional to $M\pm N$)
\begin{align}
g_{00}=& 1-G\frac{M+N}{r+a\cosh\theta+GN}-G\frac{M-N}{r-a\cosh\theta-GN}\,. \label{eq_nonlinKTNpot_selfdual}\\
\equiv &  1- \phi^+_{\textrm{non-linear}} - \phi^-_{\textrm{non-linear}} 
\end{align}
In order to check that this matches the linearized result \eqref{eq:pmphi}, we introduce the coordinates 
\begin{align}
x +i y& =\sqrt{r^{2}-a^{2}}e^{i\phi}\sinh\theta\,,\label{eq:lincoagain}\\
z & =r\cosh\theta\,\label{eq:lincoagainz}.
\end{align}
In these coordinates, the linearized potentials \eqref{eq:pmphi} become
\begin{align}
   \phi^{\pm}(x)= \frac{G(M\pm N)}{r\pm a\cosh\theta}\,\label{eq:pmphirtheta},
\end{align}
which indeed match with \eqref{eq_nonlinKTNpot_selfdual} at linear order in $G$! This is an explicit verification of the correspondence \eqref{eq:hfromM} for the Kerr-Taub-NUT metric.

\subsubsection*{Diffeomorphism from Amplitudes}

We can now revisit the diffeomorphism between self-dual Taub-NUT and Kerr-Taub-NUT found in section \ref{sec:largediffeoKTN}. As previously noted, in the $M=N$ case, $\phi^+$ and $\phi^-$ reduce to 
\begin{align}
    \phi^+ &= \frac{2G M}{\sqrt{(z+a)^2 - x^2 - y^2}} \\
    \phi^- &= 0.
\end{align}
so that the parameter $a$ can be removed by a simple shift in the $z$ coordinate. Indeed, the same conclusion can be reached from the metric itself, which from (\ref{eq:hfromM}) (using
(\ref{eq:sdm3})) simply reads

\begin{equation}
h_{\mu\nu}^{(a)}(x)=\frac{2G M }{\pi}\int d^2\lambda e^{-k\cdot(x+a)}\epsilon_{\mu}^{-}(k)\epsilon_{\nu}^{-}(k)\,,
\end{equation}
with the shift apparent in the exponent $x^\mu +a^\mu = (t,x,y,z+a)$. Importantly, the fact that $z \to z + a$ can be regarded as a transformation of $h_{\mu \nu}$ alone is dependent on the fact that it is a Killing symmetry of the flat background metric. This is because a generic coordinate transformation would change $\eta_{\mu \nu}$, meaning one would have to consider the action of the diffeomorphism the full metric $g_{\mu \nu}$.

To close this section let us comment on how the non-linear version of the diffeomorphism can be obtained. Starting from the linearized rectangular coordinates \eqref{eq:lincoagain}, \eqref{eq:lincoagainz} we can perform the simple shift $r \to r- GM$. \footnote{This shift of $r \to  r - GM$ should be distinguished from the Talbot shift $r\to r+ GM$ \cite{osti_4764870} which turns \eqref{eq:pmphirtheta} into its non-linear version \eqref{eq_nonlinKTNpot_selfdual}. The Talbot shift also leads to a Kretschmann scalar which only has terms linear in $G$, which is a feature of so-called Kerr-Schild solutions.} This shift is such that the Rindler horizon surface in rectangular coordinates, i.e.
\begin{equation}
   (z+a)^2 - x^2 - y^2 = 0\,,
\end{equation}
corresponds to the actual horizon of (Kerr-) Taub-NUT \eqref{eq_ktnhorizonlocation}, which is distinct from the singularity in the non-linear theory. In other words, the appropriate coordinate system for non-linear Taub-NUT is
\begin{align}
x +i y& =\sqrt{(r-GM)^{2}-a^{2}}e^{i\phi}\sinh\theta\,,\label{eq:linco2}\\
z & =(r-GM)\cosh\theta\,\label{eq:linco2z}.
\end{align}
Indeed, this is the coordinate system that was applied in section \ref{sec:largediffeoKTN}. 
Our ability to obtain the non-linear version in this case resonates with the well-known integrability properties of self-dual gravity \cite{Penrose:1976jq}.

At the non-linear level, using the full coordinate system, we can confirm that the transformation $z\to z + a$ precisely induces the diffeomorphism between non-linear Taub-NUT and Kerr-Taub-NUT that was observed in section \ref{sec:largediffeoKTN}.

\section{Euclidean Kerr-Taub-NUT}
While the majority of our results thus far are in Kleinian signature, the large diffeomorphism between Taub-NUT and Kerr-Taub-NUT also holds in the Euclidean case. This will follow from analytic continuation between the Kleinian and Euclidean solutions.

\subsection{Euclidean Taub-NUT}

The Euclidean Taub-NUT metric is given by
\begin{equation}\label{etn}
    ds^2_{\textrm{TN, E}} = f(r)\left(dt - 2N \cos\theta d\phi\right)^2 +\frac{dr^2}{f(r)}+(r^2 - N^2)(d\theta^2 + \sin^2\theta d\phi^2)
\end{equation}
where 
\begin{gather}
    f(r) = \frac{r^2 - 2 M r + N^2}{r^2 - N^2} = \frac{(r-r_+)(r-r_-)}{r^2 - N^2}\\
    r_\pm = M \pm \sqrt{ M^2 - N^2}
\end{gather}
Crucially, it is related to the Kleinian Kerr-Taub-NUT metric \eqref{Kktn} by the rotation $\theta \to i \theta$. Surfaces of constant $r$ are topologically 3-spheres, and $\psi=t/2N$ parameterizes the $U(1)$ fiber over a 2-sphere with the topology of the Hopf fibration.

The metric \eqref{etn} requires periodicity conditions among the Killing directions $t$ and $\phi$ in order to be smooth on the three surfaces $\theta = 0$, $\theta = \pi$, and $r = r_+$:
\begin{align}
    (t, \phi) &\sim (t + 4 \pi N, \phi + 2 \pi) \text{, from } \theta = 0 \label{etn_period1}\\
    (t, \phi) &\sim (t - 4 \pi N, \phi + 2 \pi)\text{, from } \theta = \pi \label{etn_period2}\\
    (t, \phi) &\sim (t + 4 \pi \frac{r_+^2 - N^2}{r_+ - r_-} , \phi)\text{, from } r = r_+ \text{ with } M \neq \pm N \label{etn_t_period}.
\end{align}

However, we now have three periodicity conditions for two variables. These conditions will not all be compatible unless the $t$ periodicity in \eqref{etn_t_period} equals $\pm 8 \pi N$. This will only hold for the Taub-bolt solution, $N = \pm 4 M/5$ \cite{EGUCHI_taubbolt, PAGE1978249}. Therefore, we see that the Euclidean Taub-NUT metric given in \eqref{etn} will have conical singularities for generic values of $(M,N)$. This problem does not arise in $(2,2)$ signature, which is one way that the Taub-NUT metrics are better behaved in Kleinian signature than in Euclidean signature. 

The situation changes if $M = \pm N$. In this case, the $r = r_+$ smoothness condition is found to be
\begin{equation}
    (t, \phi) \sim (t + 8 \pi N, \phi) 
\end{equation}
which is compatible with both \eqref{etn_period1} and \eqref{etn_period2} together.
The Euclidean Taub-NUT metric with $M =  N$ is self-dual. We will consider both the self-dual and the anti-self-dual ($M=-N$) cases in this section.

\subsection{Diffeomorphism to Euclidean Kerr-Taub-NUT}\label{sec:largediffeoKTNEucl}

In section \ref{sec:largediffeoKTN} we found that in Kleinian signature, the self-dual Kerr-Taub-NUT metric is diffeomorphic to the self-dual Taub-NUT metric. That is, the rotation parameter $a$ could be removed from the metric by a large diffeomorphism. Interestingly, this result also holds in Euclidean signature. In this section, we give the explicit formula for this diffeomorphism and connect this result to previous literature.

The Euclidean Kerr-Taub-NUT metric, described by three real parameters $(M,N,a)$, is 
\begin{align}\label{ektn}
    ds^2 &= \Sigma (\frac{d{r'}^2}{\Delta} + d {\theta'}^2) + \frac{\sin^2 \theta' }{\Sigma} (a dt - \rho d \phi)^2 + \frac{\Delta}{\Sigma}(dt + A d \phi)^2 \\
    \Sigma &= {r'}^2 - (N + a \cos \theta')^2 \\
    \Delta &= {r'}^2 - 2 M r' + N^2 - a^2 \\
    A &= a \sin^2 \theta' - 2 N \cos \theta' \\
    \rho &= {r'}^2 - N^2 - a^2 = \Sigma - a A.
\end{align}
Indeed, if we take $M = \pm N$ in the Euclidean Taub-NUT metric \eqref{etn} and apply the coordinate transformation (obtained by taking $\theta\to -i\theta$ in \eqref{large_diffeo_1} and \eqref{large_diffeo_2})
\begin{align}\label{large_diffeo_1_e}
    \theta &= 2 \arctan\left( \tan\left( \frac{\theta'}{2}\right) \sqrt{\frac{r' - M \mp a}{r' - M \pm a}} \right) \\
    r &= r' \pm a \cos \theta' \label{large_diffeo_2_e}
\end{align}
we recover the Euclidean Kerr-Taub-NUT metric \eqref{ektn}. 

The existence of this diffeomorphism provides a reinterpretation for a fact which has previously appeared in the literature. In their systematic study of the Euclidean Kerr-Taub-NUT metric \cite{Ghezelbash:2007kw}, Ghezelbash, Mann, and Sorkin noted that for each mass $M$, only a discrete set of choices of $(N,a)$ can yield smooth metrics \textit{except} for the case $M = \pm N$ where any value of $a$ is allowed. This was interpreted as a 1-parameter family of exceptional solutions. Here, we have shown that the solutions in this 1-parameter family are diffeomorphic, although the diffeomorphisms are likely non-trivial. 

This fact also gives a natural interpretation to the Newman-Janis trick in Euclidean signature. One can very schematically understand a Euclidean Schwarzschild black hole as a point of mass and NUT charge $(M,N)$ superimposed over a point of mass and NUT charge $(M,-N)$. Introducing the rotation parameter $a$ into the metric translates the $(M,N)$ mass in the $z$-direction by $a$ and the $(M,-N)$ mass by $-a$. Therefore, the Euclidean Kerr black hole can itself be understood as a $(M,N)$ point mass separated from a $(M,-N)$ point mass by a distance of $2a$. This was demonstrated in the classic paper of Gross and Perry \cite{Gross:1983hb} which showed that the self-dual Euclidean Taub-NUT metric can be understood as the spatial part of a Kaluza-Klein magnetic monopole in 4+1 dimensions, while the Euclidean Kerr metric can be understood as a magnetic dipole with one monopole and one anti-monopole separated by a distance of $2a$. In contrast, our Kerr-Taub-NUT diffeomorphism can be understood as starting with only a single point mass of $(M,N)$ at the origin, with no partner mass, and translating it by $a$.

\section{Discussion}\label{sec:dscs}

In this work we have constructed the $(2,2)$ versions of self-dual Taub-NUT and Schwarzschild spacetimes. Extending the analysis of \cite{Atanasov:2021celesttorus} for flat space, we have found the global structure corresponding to the maximal extension of these spacetimes. Remarkably, null infinity was found to consist of a single connected component. We expect this to have implications for constructing the associated ``S-vector'' for the scattering problem in such backgrounds.

For self-dual Taub-NUT, we found that the spin parameter $a$ can be added to the solution through a diffeomorphism $z\to z + a$. The result is a $(2,2)$ metric that agrees with the analytic continuation of the known Kerr-Taub-NUT solution. The transformation is reminiscent of the Newman-Janis shift, but has crucial differences: it only applies to the (anti-) self-dual metrics, and does not require a special procedure to preserve the reality of the metric since it is a bona fide coordinate change.

Both the analytic continuation and the diffeomorphism are directly linked to underlying scattering amplitudes. We found that analyticity is a powerful tool even at the linearized level, where we can treat stationary metrics as perturbations of Klein space. Imposing that the linearized metric is well-defined in momentum space restricts the perturbations to live on a Rindler wedge of Klein space, bounded by a singularity. In this wedge, the linearized field is free and can be connected to scattering amplitudes.

In a Lorentzian spacetime, a field is assumed to be free only in the asymptotic regions. Stationary black hole solutions do not contain radiative data associated to gravitons piercing null infinity. Because of this, previous work to find connections between scattering amplitudes and the Kerr/Taub-NUT/Plebanski metrics \cite{Cachazo:2017jef,Kosower:2018adc,Arkani-Hamed:2019ymq, Emond:2020lwi, Guevara:2018wpp,Chung:2018kqs,Huang:2019cja} proceeded instead by considering a system of \textit{two} weakly coupled black holes, or a black hole and a test particle, which can interact through graviton exchange. These amplitudes could then be linked to three-point amplitudes using factorization. In contrast to this, \cite{Monteiro:2020plf} extended the previous work of \cite{Kosower:2018adc} to more directly study three-point amplitudes in $(2,2)$ signature. It was found that the microscopic graviton states can be exponentiated to produce a single classical on-shell solution without the need to go to null infinity. Our discussion is consistent with their findings: we showed that only on-shell modes can appear in the stationary $(2,2)$ solution. 

Indeed, we can further comment on the relation between our results and those of \cite{Monteiro:2020plf}. The authors explicitly constructed a new $(2,2)$ metric, analogous to the Schwarzschild metric, from three-point amplitudes. By construction, it is composed of purely free modes and it is defined in the full Klein space. Then, the classical derivation of the metric is provided by matching the new metric to a source at the origin, which requires the use of a $i\epsilon$-prescription corresponding to retarded boundary conditions. Due to this prescription the metric exhibits certain discontinuities. On the other hand, here we have taken a somewhat opposite route and started from the analytic continuation of classical Lorentzian metrics. In contrast to \cite{Monteiro:2020plf}, analytic continuation leads to smooth $(2,2)$ metrics by construction, but they are only defined on a wedge region of Klein space.
In our results, the fact that the linearized field becomes free in this wedge leads to a natural connection with scattering amplitudes. Now, analytic continuation also implies that our contour prescription in momentum space is equivalent to the Feynman prescription rather than the retarded boundary condition. This simply means that in the contour integral of section \ref{subsec:sec-cont-lin} we picked one pole rather than both.
This yields an overall factor of 2 when the Schwarzschild metric of \cite{Monteiro:2020plf} is restricted to our wedge region. We confirm this when we study the Schwarzschild case in Appendix \ref{sec:app_schwarz} via the continuation procedure.

Finally, a relation between the Newman-Janis shift and spinning amplitudes was previously observed in \cite{Arkani-Hamed:2019ymq}. Other relations between Schwarzschild, Taub-NUT and Kerr-Taub-NUT metrics have since been further understood \cite{Banerjee:2019saj,Alawadhi:2019urr,Huang:2019cja,Moynihan:2020gxj}. The well-known double-copy structure of gravity amplitudes in terms of gauge-theory amplitudes has motivated the interpretation of these classical spaces as double copies of classical dyons \cite{Luna:2015paa,Alfonsi:2020lub,Bahjat-Abbas:2020cyb,Emond:2021lfy,Csaki:2020inw}. On the other hand, in contrast to previous approaches, here we have provided a direct connection between stationary metrics and three-point amplitudes. Revisiting the above work with our direct connection in mind could provide interesting future directions.

\subsection*{Acknowledgements}

We thank Adam Ball and Justin Vines for useful discussions. This work was supported by DOE grant de-sc/000787. AG is supported by the Black Hole
Initiative at Harvard University, which is funded by grants from the John Templeton Foundation and the Gordon and Betty Moore Foundation and  the Harvard Society of Fellows. NM gratefully acknowledges support from NSF GRFP grant
DGE1745303. EC acknowledges support from the Ashford Fellowship at Harvard University.

\appendix

\section{Symmetries of the Kleinian Taub-NUT metric} \label{sec:app_TNsymmetries}

The $(2,2)$ Taub-NUT metric \eqref{eq_22TaubNUTgeneric} has a four-dimensional group of Killing symmetries which is $\mathfrak{sl}(2, \mathbb{R}) \times \mathfrak{u}(1)$. This is analytically continued from the  $\mathfrak{su}(2) \times \mathfrak{u}(1) $ symmetry of the Euclidean Taub-NUT metric. After defining $\psi = t / 2N$, the Killing vectors are given by
\begin{align}
    \xi_{1} &= \frac{\cos \phi}{\sinh \theta} \partial_\psi + \sin \phi \partial_\theta + \cos \phi \coth \theta \partial_\phi \\
    \xi_{2} &= -\frac{\sin \phi}{\sinh \theta} \partial_\psi + \cos \phi \partial_\theta - \sin \phi \coth \theta \partial_\phi \\
    \xi_{3} &= - \partial_\phi \\
    \xi_{4} &= \partial_\psi
\end{align}
which satisfy the commutation relations
\begin{align}
    [\xi_{1},\xi_{2}] &=\xi_{3} \\
    [\xi_{2},\xi_{3}] &=-\xi_{1} \\
    [\xi_{3},\xi_{1}] &=-\xi_{2} \\
    [ \xi_{4},\xi_{i}] &= 0 \hspace{1 cm} i = 1,2,3
\end{align}
Using the 1-forms
\begin{align}
    \sigma_{1} &= \tfrac{1}{2}( -\sin \psi d \theta - \cos \psi \sinh\theta d \phi) \\
    \sigma_{2} &= \tfrac{1}{2}( \cos \psi d \theta - \sin \psi \sinh\theta d \phi) \\
    \sigma_{3} &= \tfrac{1}{2}( d \psi - \cosh \theta d \phi)
\end{align}
we can rewrite the metric \eqref{eq_22TaubNUTgeneric} as
\begin{equation}\label{metric_1form}
    ds^2 = 4 N^2 f(r) \sigma_{3}^2 + \frac{dr^2}{f(r)} - 4(r^2 - N^2) (\sigma_{1}^2 + \sigma_{2}^2).
\end{equation}
For $i, j = 1, 2, 3$, the Lie derivatives of the Killing vector $\xi_{i}$ annihilate $\sigma_{j}$.
\begin{equation}\label{lie_xi_sigma1}
    \mathcal{L}_{\xi_{i}} \sigma_{j} = 0.
\end{equation}
This is because the 1-forms $\sigma_i$ correspond to the right-invariant 1-forms on the $SL(2,\mathbb{R})$ group manifold, while the Killing vectors $\xi_i$ generate the right action on the manifold.

Furthermore, $\xi_{4}$ annihilates $\sigma_{3}$ as well as the combination $\sigma_{1}^2 + \sigma_{2}^2$.
\begin{equation}\label{lie_xi_sigma2}
    \mathcal{L}_{\xi_{4}} \sigma_{3} = 0 \hspace{1.5 cm} \mathcal{L}_{\xi_{4}} ( \sigma_{1}^2 + \sigma_{2}^2 )= 0.
\end{equation}
From \eqref{metric_1form}, \eqref{lie_xi_sigma1}, and \eqref{lie_xi_sigma2}, one can immediately see that $\mathcal{L}_{\xi} g_{\mu \nu} = 0$ for all four Killing vectors using the product rule of the Lie derivative.

In the form of the metric \eqref{metric_1form}, it is manifest that each $r = const$ surface is a warped $AdS_3$ manifold \label{Bengtsson:2005zj} because the combination $\sigma_1^2 + \sigma_2^2 - \sigma_3^2$ is equal to the unwarped $AdS_3$ metric.

\section{Schwarzschild and \texorpdfstring{$M=0$}{M=0} Taub-NUT Toric Penrose Diagrams} \label{sec:app_schwarz}

In this section, we will consider two other useful spacetimes in Kleinian signature, where either the mass or the NUT charge in the Taub-NUT metric \eqref{eq_22TaubNUTgeneric} is taken to be zero.

\subsection{Kleinian Schwarzschild}
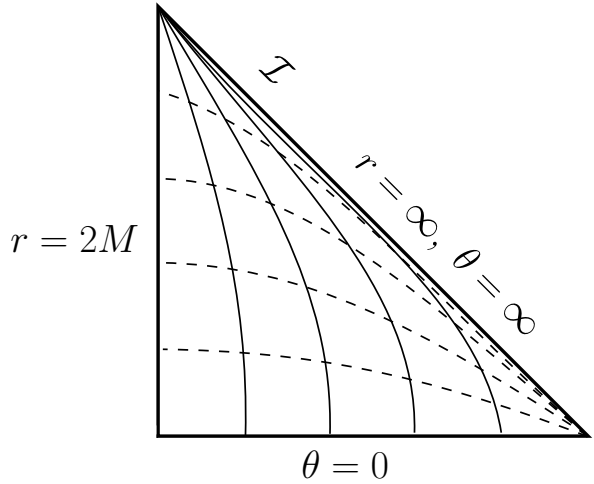
\begin{figure}[ht]
    \centering
    \resizebox{0.5\textwidth}{!}    {
        \tikzset{every picture/.style={line width=0.75pt}} 

\begin{tikzpicture}[x=0.75pt,y=0.75pt,yscale=-1,xscale=1]

\draw  [color={rgb, 255:red, 0; green, 0; blue, 0 }  ,draw opacity=1 ][line width=1.5]  (228.67,16.2) -- (479.67,267.2) -- (228.67,267.2) -- cycle ;
\draw    (228.67,16.2) .. controls (262.67,117.2) and (281.67,183.2) .. (279.67,267.2) ;
\draw    (228.67,16.2) .. controls (289,117.2) and (331,181.87) .. (329,265.87) ;
\draw    (228.67,16.2) .. controls (307,115.87) and (380.33,181.2) .. (378.33,265.2) ;
\draw    (228.67,16.2) .. controls (315.67,121.87) and (415.67,180.53) .. (429,265.2) ;
\draw  [dash pattern={on 4.5pt off 4.5pt}]  (479.53,267.47) .. controls (379.97,229.46) and (315.45,217.84) .. (231.44,216.48) ;
\draw  [dash pattern={on 4.5pt off 4.5pt}]  (479.67,267.2) .. controls (381.16,202.88) and (313.42,167.26) .. (229.41,165.91) ;
\draw  [dash pattern={on 4.5pt off 4.5pt}]  (479.53,267.47) .. controls (383.07,185.21) and (312.35,118.16) .. (228.33,116.8) ;
\draw  [dash pattern={on 4.5pt off 4.5pt}]  (479.53,267.47) .. controls (377.43,176.32) and (315.67,92.13) .. (228.33,65.47) ;
\draw [line width=1.5]    (222.78,267.24) -- (228.67,267.2) ;
\draw  [color={rgb, 255:red, 255; green, 255; blue, 255 }  ,draw opacity=1 ][fill={rgb, 255:red, 255; green, 255; blue, 255 }  ,fill opacity=1 ] (210.65,261.47) -- (227.13,261.47) -- (227.13,281.47) -- (210.65,281.47) -- cycle ;

\draw (180.03,151.6) node  [font=\LARGE,color={rgb, 255:red, 0; green, 0; blue, 0 }  ,opacity=1 ]  {$\textcolor[rgb]{0,0,0}{r=2M}$};
\draw (370,128) node  [font=\LARGE,rotate=-45]  {$\mathcal{I}$\quad\quad\quad $r=\infty ,\ \theta =\infty $};
\draw (337.86,283.52) node  [font=\LARGE,rotate=-360]  {$\theta =0$};

\end{tikzpicture}
    }
    \caption{The toric Penrose diagram for $(2,2)$ Schwarzschild. Lines of constant $r$ are solid, and lines of constant $\theta$ are dashed. }
    \label{fig_schwarzschild}
\end{figure}

The Kleinian Schwarzschild metric can be obtained by setting $N=0$ in the generic Taub-NUT metric \eqref{eq_22TaubNUTgeneric}. It takes the form
\begin{equation}
    ds_{\rm Sch}^2 = \left(1-\frac{2M}{r}\right)dt^2 +\frac{dr^2}{1-\frac{2M}{r}} - r^2(d\theta^2 + \sinh^2\theta d\phi^2), \label{eq_kleinschwarz}
\end{equation}
where $t \sim t+8\pi M$ and $\phi \sim \phi + 2\pi$. This metric has the geometry of a spacelike ``cigar'' in $t$ and $r$ crossed with a timelike hyperbolic plane in $\theta$ and $\phi$. So, we have $r\in [2m, \infty)$ and $\theta\in[0, \infty)$. Moreover, since each of the cigar and the hyperbolic plane have topology $\mathbb{R}^2$, this metric has global topology $\mathbb{R}^4$. Since $\phi$ and $t$ are compact directions, we can set $d\phi=dt = 0$ to find the toric Penrose diagram in $r$ and $\theta$. This yields the $(1,1)$ metric
\begin{gather}
    ds_{\rm Sch, 2d}^2 = \frac{dr^2}{ 1-\frac{2M}{r}} -  r^2 d\theta^2. \label{eq_TN_Klein_2d_schwarz}
\end{gather}
Considering this metric for large $r$ yields
\begin{equation}
        ds_{\rm Sch, 2d}^2 \approx dr^2 - r^2 d\theta^2, 
\end{equation}
which is the Rindler metric for flat space. So, null infinity $\mathcal{I}$ can be reached. Morever, the spacetime is geodesically complete. We can also consider $r$ near $2M$ by letting $r = 2M + \frac{\rho^2}{2M}$, obtaining
\begin{equation}
    ds^2_{\textrm{Sch,2d}} \approx 4d\rho^2 - 4M^2d\theta^2.
\end{equation} 
After rescaling $\rho$ and $\theta$, this is $(1,1)$ Minkowski space. The spacetime can be conformally compactified as in \cite{Atanasov:2021celesttorus}, and the toric Penrose diagram for $(2,2)$ Schwarzschild is shown in Fig. \ref{fig_schwarzschild}.

We also stop here to comment on the Kleinian Schwarzschild solution recently found in \cite{Monteiro:2020plf}, and show its agreement with our results in a certain region (up to a factor of 2 from differing prescriptions, as mentioned in the discussion). The Kleinian Schwarzschild metric from \cite{Monteiro:2020plf} is given in static coordinates $(t,x,y,z)$ as
\begin{equation}
    ds^2 = (1- \Phi) dt^2 + dz^2 - dx^2 - dy^2 + \frac{\Phi}{1-\Phi} \frac{(z dz - xdx - ydy)^2}{z^2 - x^2 - y^2},
\end{equation}
where the gravitational potential is given by
\begin{equation}
    \Phi = 4 G M \frac{\Theta(z-\sqrt{x^2 + y^2})}{\sqrt{z^2 - x^2 - y^2}}. \label{eq_monteirograv}
\end{equation}
We note that this potential is smooth in the Rindler wedge. This is precisely the region of convergence for the  Schwarzschild/Taub-NUT potential $\Phi=-\phi^+ - \phi^-$ derived in section \ref{sec:linfromamp}. The gravitational potential \eqref{eq_monteirograv} agrees with our result up to a factor of 2, as previously mentioned.


Once again, for the Schwarzschild case one can check that rotating $\theta\to i\theta$ and $t \to i t$ is equivalent to sending $(t,x,y)\to (it, ix, iy)$ in rectangular coordinates, as long as we are inside the Rindler wedge.

\subsection{Kleinian $M=0$ Taub-NUT}
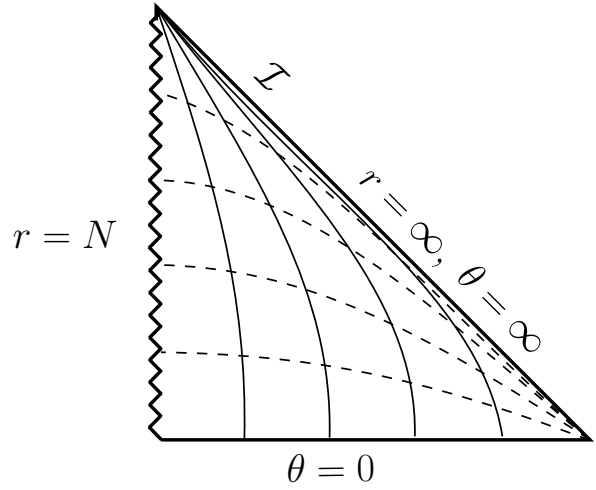
\begin{figure}[ht]
    \centering
    \resizebox{0.5\textwidth}{!}    {
        \tikzset{every picture/.style={line width=0.75pt}} 

\begin{tikzpicture}[x=0.75pt,y=0.75pt,yscale=-1,xscale=1]

\draw  [color={rgb, 255:red, 0; green, 0; blue, 0 }  ,draw opacity=1 ][line width=1.5]  (228.67,16.2) -- (479.67,267.2) -- (228.67,267.2) -- cycle ;
\draw    (228.67,16.2) .. controls (262.67,117.2) and (281.67,183.2) .. (279.67,267.2) ;
\draw    (228.67,16.2) .. controls (289,117.2) and (331,181.87) .. (329,265.87) ;
\draw    (228.67,16.2) .. controls (307,115.87) and (380.33,181.2) .. (378.33,265.2) ;
\draw    (228.67,16.2) .. controls (315.67,121.87) and (415.67,180.53) .. (429,265.2) ;
\draw  [dash pattern={on 4.5pt off 4.5pt}]  (479.53,267.47) .. controls (379.97,229.46) and (315.45,217.84) .. (231.44,216.48) ;
\draw  [dash pattern={on 4.5pt off 4.5pt}]  (479.67,267.2) .. controls (381.16,202.88) and (313.42,167.26) .. (229.41,165.91) ;
\draw  [dash pattern={on 4.5pt off 4.5pt}]  (479.53,267.47) .. controls (383.07,185.21) and (312.35,118.16) .. (228.33,116.8) ;
\draw  [dash pattern={on 4.5pt off 4.5pt}]  (479.53,267.47) .. controls (377.43,176.32) and (315.67,92.13) .. (228.33,65.47) ;
\draw [line width=1.5]    (222.78,267.24) -- (228.67,267.2) ;
\draw  [color={rgb, 255:red, 255; green, 255; blue, 255 }  ,draw opacity=1 ][fill={rgb, 255:red, 255; green, 255; blue, 255 }  ,fill opacity=1 ] (220.5,257.8) -- (234,271.3) -- (222,283.3) -- (208.5,269.8) -- cycle ;
\draw  [color={rgb, 255:red, 255; green, 255; blue, 255 }  ,draw opacity=1 ][fill={rgb, 255:red, 255; green, 255; blue, 255 }  ,fill opacity=1 ] (225.16,24.8) -- (229.56,24.8) -- (229.56,264.88) -- (225.16,264.88) -- cycle ;
\draw [line width=1.5]    (224.68,207.65) -- (231.68,214.77) ;
\draw [line width=1.5]    (224.68,208.16) -- (231.68,201.03) ;
\draw [line width=1.5]    (224.68,194.76) -- (231.68,201.88) ;
\draw [line width=1.5]    (224.68,195.27) -- (231.68,188.14) ;
\draw [line width=1.5]    (224.68,181.66) -- (231.68,188.79) ;
\draw [line width=1.5]    (224.68,182.17) -- (231.68,175.05) ;
\draw [line width=1.5]    (224.68,168.77) -- (231.68,175.9) ;
\draw [line width=1.5]    (224.68,169.28) -- (231.68,162.16) ;
\draw [line width=1.5]    (224.48,259.76) -- (231.48,266.88) ;
\draw [line width=1.5]    (224.48,260.26) -- (231.48,253.14) ;
\draw [line width=1.5]    (224.68,246.66) -- (231.68,253.79) ;
\draw [line width=1.5]    (224.68,247.17) -- (231.68,240.05) ;
\draw [line width=1.5]    (224.68,233.57) -- (231.68,240.69) ;
\draw [line width=1.5]    (224.68,234.08) -- (231.68,226.95) ;
\draw [line width=1.5]    (224.68,220.68) -- (231.68,227.8) ;
\draw [line width=1.5]    (224.68,221.18) -- (231.68,214.06) ;
\draw [line width=1.5]    (224.48,129.49) -- (231.48,136.61) ;
\draw [line width=1.5]    (224.48,130) -- (231.48,122.87) ;
\draw [line width=1.5]    (224.48,116.6) -- (231.48,123.72) ;
\draw [line width=1.5]    (224.48,117.11) -- (231.48,109.98) ;
\draw [line width=1.5]    (224.48,103.5) -- (231.48,110.63) ;
\draw [line width=1.5]    (224.48,104.01) -- (231.48,96.89) ;
\draw [line width=1.5]    (224.48,155.61) -- (231.48,162.73) ;
\draw [line width=1.5]    (224.48,156.12) -- (231.48,149) ;
\draw [line width=1.5]    (224.48,142.52) -- (231.48,149.64) ;
\draw [line width=1.5]    (224.48,143.02) -- (231.48,135.9) ;

\draw [line width=1.5]    (225.01,38.31) -- (232.02,45.43) ;
\draw [line width=1.5]    (225.01,38.81) -- (232.02,31.69) ;
\draw [line width=1.5]    (225.01,25.41) -- (232.02,32.54) ;
\draw [line width=1.5]    (225.01,25.92) -- (232.02,18.8) ;
\draw [line width=1.5]    (224.81,90.41) -- (231.81,97.54) ;
\draw [line width=1.5]    (224.81,90.92) -- (231.81,83.8) ;
\draw [line width=1.5]    (225.01,77.32) -- (232.02,84.44) ;
\draw [line width=1.5]    (225.01,77.83) -- (232.02,70.7) ;
\draw [line width=1.5]    (225.01,64.22) -- (232.02,71.35) ;
\draw [line width=1.5]    (225.01,64.73) -- (232.02,57.61) ;
\draw [line width=1.5]    (225.01,51.33) -- (232.02,58.46) ;
\draw [line width=1.5]    (225.01,51.84) -- (232.02,44.72) ;

\draw (175.98,148) node  [font=\LARGE,color={rgb, 255:red, 0; green, 0; blue, 0 }  ,opacity=1 ]  {$\textcolor[rgb]{0,0,0}{r=N}$};
\draw (329.76,283.45) node  [font=\LARGE,rotate=-359.24]  {$\theta =0$};
\draw (370,134.11) node  [font=\LARGE,rotate=-45]  {$\mathcal{I}$ \quad\quad\quad$r=\infty ,\ \theta =\infty $};

\end{tikzpicture}
    }
    \caption{The toric Penrose diagram for $(2,2)$ Taub-NUT with $M=0$. There is a naked singularity at $r=N$. Lines of constant $r$ are solid, and lines of constant $\theta$ are dashed.}
    \label{fig_Mzero}
\end{figure}

Now, we consider the case of Taub-NUT with $M=0$. We will find that it does not exactly parallel the Schwarzschild case, and instead has a naked singularity. The metric is 
\begin{equation}
    ds^2 = \frac{r^2 + N^2}{r^2 - N^2}(dt - 2 N \cosh \theta d \phi)^2 + \frac{r^2 - N^2}{r^2 + N^2}dr^2 - (r^2 - N^2)( d \theta^2 + \sinh^2 \theta d \phi^2 ),
\end{equation}
where $t$ and $\phi$ have periodicity as given in \eqref{eq_22TN_MnotN_period1} and \eqref{eq_22TN_MnotN_period2}. The Kretschmann scalar for this metric is
\begin{equation}
 R_{\mu \nu \rho \sigma} R^{\mu \nu \rho \sigma} = 24N^2 \left(\frac{1}{(N - r)^6} + \frac{1}{(N + r)^6}\right),
\end{equation}
so there are possible singularities at $r=\pm N$. As in the previous cases, we let $d\phi= dt=0$, yielding the $(1,1)$ metric
\begin{equation}
    ds^2_{\textrm{2d}} = \frac{r^2 - N^2}{r^2 + N^2}dr^2 - (r^2 - N^2)d \theta^2. 
\end{equation}
For large $r$, this is 
\begin{equation}
    ds^2_{\textrm{2d}} \approx {dr^2} - r^2 d\theta^2\,,
\end{equation}
which is again flat (Rindler) space, reaching null infinity. At $r=N$, there is a naked singularity. The toric Penrose diagram is shown in Fig. \ref{fig_Mzero}.\\

\section{Torus Modular Parameter}\label{app_torus}

In this appendix, we will derive the modular parameter on the torus for self-dual Klein Taub-NUT found in \eqref{eq_torusmodparam}. 
We will show how to find a generic modular parameter by first starting with a general 2d metric
\begin{align}
    ds^2 &= a \, dx^2 + 2 b \, dx dy + c \, dy^2,
\end{align}
where the two real variables $x, y$ have periodicity conditions
\begin{align}
    (x,y) &\sim (x + 2 \pi, y), \\
    (x,y) &\sim (x, y + 2 \pi).
\end{align}
We further assume the above metric has signature $(1,1)$. We perform a change of coordinates so that the metric becomes 
\begin{equation}
    ds^2 = - d{x'}^2 + d{y'}^2
\end{equation} and the periodicity conditions are
\begin{align}
    (x', y') &\sim (x' + 2 \pi, y') \\
    (x', y') &\sim (x' + 2 \pi \mathrm{Re}(\tau) , y' + 2 \pi \mathrm{Im}(\tau) ) 
\end{align}
where $\tau$ is the modular parameter of this torus. This is found to be
\begin{equation}\label{tau_abc}
    \tau = \frac{b}{a} + i \sqrt{\left( -\frac{c}{a} + \frac{b^2}{a^2} \right)}.
\end{equation}
Now, we specialize to the case of self-dual Taub-NUT. We begin with the warped $AdS_3$ metric
\begin{equation}
    ds^2 = F ( d \psi - \cosh \theta d \phi)^2 - (d \theta^2 + \sinh^2 \theta d \phi^2 )
\end{equation}
where $F$ is the warping factor. If we define the variables
\begin{equation}
    \phi' = \frac{\phi + \psi }{2} \hspace{1 cm}
    \psi' = \frac{\phi - \psi}{2} 
\end{equation}
which have the periodicities
\begin{align}
    (\phi,\psi) &\sim (\phi + 2 \pi, \psi) \\
    (\phi,\psi) &\sim (\phi, \psi + 2 \pi)
\end{align}
then, after writing the metric in $(\theta, \phi', \psi')$ coordinates, it turns out that there are no $d \theta d \phi'$ or $d \theta d \psi'$ cross terms. Applying \eqref{tau_abc} to the $(\phi', \psi')$ part of the metric yields the result
\begin{equation}
    \tau = \left(1 - \frac{2F}{F + 1 + (F-1) \cosh \theta} \right) + i \frac{2 \sqrt{F}  \tanh(\theta/2) }{F + 1 + (F - 1) \cosh \theta}.
\end{equation}

\section{Connection Between Radiative Solutions and Amplitudes}\label{ap:radamp}

In this appendix we outline the correspondence between radiative spacetimes and the classical limit of massive scattering amplitudes that has recently emerged in $(1,3)$ signature. The derivation essentially follows that of LSZ reduction, except that we will scatter
classical wavepackets as in \cite{Kosower:2018adc}. 

First, we will outline how the radiative field, \eqref{eq:hradT}, is defined as the difference between retarded and advanced propagators.\footnote{As this discussion is entirely in Lorentzian signature we will omit the superscript \textrm{`L'}.} We recall from \eqref{eq_lorEinsteineqn} and \eqref{eq_lorhwithretprop} that the Einstein equation takes the form 
\begin{equation}
\partial^{2}\bar{h}_{\mu\nu}=-16\pi G\mathcal{T}_{\mu\nu}(x)\,,
\end{equation}
where one can solve for $\bar{h}_{\mu \nu}$ as 
\begin{equation}
\bar{h}_{\mu\nu}=16\pi G\int d^4 y G_{\rm ret}(x-y)\mathcal{T}_{\mu\nu}(y)\,,\label{eq_applorhwithretprop}
\end{equation}
using the retarded propagator
\begin{equation}
G_{\rm ret}(x)= \int\frac{d^{4}k}{(2\pi)^{4}}\frac{  e^{ik\cdot x}}{(k^0 + i\epsilon)^{2} - \vec{k}^2} = \frac{\delta(x^0 - |\vec{x}|)}{2\pi |\vec{x}|} .
\end{equation}
One can define the ``radiative'' propagator as the difference between the retarded and advanced propagators (where the advanced propagator $G_{\rm adv}(x-y)$ is defined by the other $i\epsilon$ prescription)
\begin{align}
    G_{\rm rad}(x-y) =& \,\, G_{\rm ret}(x-y) -G_{\rm adv}(x-y)\\
                    =& \int\frac{d^{4}k}{(2\pi)^{3}}i \, \mathrm{sign}(k^0)\delta(k^2) e^{i k\cdot (x-y)}. 
\end{align}
Since $G_{\rm rad}(x-y)$ only has support for on-shell ($k^2 = 0$) Fourier modes, it solves the wave equation $\partial^2 G_{\rm rad}=0$. Moreover, we can rewrite \eqref{eq_applorhwithretprop} as
\begin{equation}
\bar{h}_{\mu\nu}=16\pi G\int d^4 y \left(G_{\rm rad}(x-y) + G_{\rm adv}(x-y)\right)\mathcal{T}_{\mu\nu}(y)\,.
\end{equation}
We note that $G_{\rm adv}(x-y)$ has support only when $x$ is in the past lightcone of $y$. Therefore, if we are only concerned with the metric perturbations which reach future null infinity $\mathcal{I}^+$, the advanced propagator drops out of the expression. We can accordingly define the radiative piece of the perturbation as
\begin{align}
\bar{h}^{\rm rad}_{\mu\nu}=&\,\,16\pi G\int d^4 y G_{\rm rad}(x-y)\mathcal{T}_{\mu\nu}(y)\nonumber\,\\
                        =& \,\, 16\pi G \int\frac{d^{4}k}{(2\pi)^{3}} \, i\mathrm{sign}(k^0)\delta(k^2)  e^{i k\cdot x}   \mathcal{T}_{\mu\nu}(k)\,.
\end{align}
This concludes the derivation of \eqref{eq:hradT}.

Now, we move on to outlining the connection between these radiative spacetimes and classical massive scattering amplitudes. We can write \eqref{eq:hradT} as
\begin{align}
\bar{h}_{\mu\nu}^{\text{rad}} & \equiv16\pi G\int\frac{d^{4}k}{(2\pi)^{3}}\,i\mathrm{sign}(k^0)\delta(k^{2})e^{ik\cdot x}\mathcal{T}_{\mu\nu}(k)\,,\nonumber \\
 & =16\pi G\int\frac{d^{4}k}{(2\pi)^{3}}\,i\delta(k^{2})\theta(k^{0})\left[e^{ik\cdot x}\mathcal{T}_{\mu\nu}(k)-e^{-ik\cdot x}\mathcal{T}_{\mu\nu}(-k)\right]\nonumber \\
 & =-16\pi G\int\frac{d^{3}k}{(2\pi)^{3}}\,\frac{1}{|\vec{k}|}\textrm{Im}\left[e^{ik\cdot x}\mathcal{T}_{\mu\nu}(k)\right]
\end{align}
In the last line we have used that $\mathcal{T}_{\mu\nu}(-k)=\mathcal{T}_{\mu\nu}^{*}(k)$, which follows the fact that $h_{\mu\nu}(x)$ in \eqref{eq_lorhwithretprop} is a real-valued function. As $\mathcal{T}_{\mu\nu}$
is evaluated at $k^{2}=0$ (or equivalently, the field $\bar{h}_{\mu\nu}^{\text{rad}}$
is free), we can expand into two polarizations 

\begin{equation}
\mathcal{T}_{\mu\nu}(k^{2}=0)=i\epsilon_{\mu\nu}^{\alpha}(k)a_{\alpha}(k)\,\,,\quad\alpha=\pm\,.\label{eq:tintoa}
\end{equation}
Then, we have

\begin{align}
\bar{h}_{\mu\nu}^{\text{rad}} & =-16\pi G\int\frac{d^{3}k}{(2\pi)^{3}}\,\frac{1}{2|\vec{k}|}\left[e^{ik\cdot x}\epsilon_{\mu\nu}^{\alpha}a_{\alpha}(k)+e^{-ik\cdot x}\epsilon_{\mu\nu}^{\alpha*}a_{\alpha}^{\dagger}(k)\right]\\
\stackrel{r\to\infty}{\longrightarrow}\bar{h}_{\mu\nu}^{\textrm{out }} & =16\pi G\times\frac{i}{8\pi^{2}r}\int_{0}^{\infty}d\omega\,\left[e^{i\omega u}\epsilon_{\mu\nu}^{\alpha}(\omega\hat{x})a_{a}(\omega\hat{x})-e^{-i\omega u}\epsilon_{\mu\nu}^{\alpha*}(\omega\hat{x})(a_{\alpha})^{\dagger}(\omega\hat{x})\right]\,,
\end{align}
where in the last line we have used the usual asymptotics for the
mode expansion as $r\to\infty$, for $u=t-r$ fixed \cite{Strominger:2017zoo}. We now
regard $\bar{h}_{\mu\nu}^{\text{out }}$ as a quantum operator inserting
outgoing graviton states. 

We want to consider the expectation value of the radiation field sourced
by massive particles. Let us denote the early-time wavefunction of
the massive particles by $|\psi\rangle$; following \cite{Cristofoli:2021vyo}, this wavefunction
is very localized, so we can assume the particles are classical. As there
is no radiation at early times we have $\langle\psi|\bar{h}_{\mu\nu}^{\textrm{out }}|\psi\rangle=0$.
At late times, we have the expectation value
\begin{align*}
\langle\bar{h}_{\mu\nu}^{\text{out }}\rangle & \equiv\langle\psi|S^{\dagger}\bar{h}_{\mu\nu}^{\textrm{out }}S|\psi\rangle\\
 & \approx\langle\psi|\bar{h}_{\mu\nu}^{\textrm{out }}S|\psi\rangle+\langle\psi|S^{\dagger}\bar{h}_{\mu\nu}^{\textrm{out }}|\psi\rangle
\end{align*}
Where $\approx$ means we have ignored quadratic terms in the coupling. Inserting the asymptotic expansion, we obtain
\begin{equation}
\langle\bar{h}_{\mu\nu}^{\text{out }}\rangle=\frac{16\pi G }{8\pi^{2}r}i\int_{0}^{\infty}d\omega\,\left[e^{i\omega u}\epsilon_{\mu\nu}^{\alpha}(\omega\hat{x})\langle\psi|a_{a}(\omega\hat{x})S|\psi\rangle-e^{-i\omega u}\epsilon_{\mu\nu}^{\alpha*}(\omega\hat{x})\langle\psi|S^{\dagger}(a_{\alpha})^{\dagger}(\omega\hat{x})|\psi\rangle\right]
\end{equation}

Thus we have found that we can compute the expectation value of the
outgoing field simply by replacing the radiative degrees of freedom
as

\begin{equation}
a_{\pm}\to\langle\psi|a_{\pm}(\omega\hat{x})S|\psi\rangle=i\mathcal{M}_{\pm}(\psi,k)
\end{equation}
which is a scattering amplitude for a massive source and a graviton of
momentum $k^\mu=(\omega,\omega\hat{x})$. These are the so-called Newman-Penrose amplitudes recently studied in \cite{Bautista:2021wfy,Cristofoli:2021vyo}. From (\ref{eq:tintoa}) this yields
\begin{equation}
\epsilon_{\pm}^{\mu\nu}(k)\mathcal{T}_{\mu\nu}(k)\rightarrow\mathcal{M}_{\pm}(\psi,k)
\end{equation}
as would be expected from the LSZ formula. We should note that the precise equality between the two (i.e. adopting the classical limit not only of the massive source but also the massless particle), requires one to consider a coherent superposition of infinite graviton states, generating a macroscopic field as in \cite{Cristofoli:2021vyo}.

\section{Spinorial Version and Curvature}\label{app_spinor}

In this appendix, we will compute the linearized Weyl curvature tensor associated to the metric perturbation $h_{\mu\nu}=h^+_{\mu\nu}+h^-_{\mu\nu}$ where
\begin{equation}\label{eq:resfr}
h_{\mu\nu}^{\pm}(x)=\frac{8\pi G}{(2\pi)^{2}}\int d^2\lambda e^{-k\cdot x}\epsilon_{\mu}^{\mp}\epsilon_{\nu}^{\mp}\mathcal{M}_{3}^{\pm}(k)\,,
\end{equation}
and show that it can be naturally written in terms of spinors. 

We start by introducing spinor conventions in $(2,2)$ signature with metric $(+,-,-,+)$. We use the convention
\begin{equation}
    \epsilon^{12} = -\epsilon_{12} = \epsilon^{\dot{1} \dot{2}} = -\epsilon_{\dot{1} \dot{2}} = 1.
\end{equation}
so that
\begin{equation}
    \epsilon_{AB} \epsilon^{BC} = \delta^C_A.
\end{equation}
We raise and lower spinors from the left as
\begin{equation}
    \lambda_A = \epsilon_{AB} \lambda^{B} = \epsilon_{AB} (\epsilon^{BC} \lambda_C)
\end{equation}
and use the inner product notation, defined by
\begin{equation}
    | \lambda \rangle \leftrightarrow \lambda_A \hspace{1 cm} \langle \lambda | \leftrightarrow \lambda^A \hspace{1 cm} | \lambda ] \leftrightarrow \widetilde{\lambda}^{\dA} \hspace{1 cm} [ \lambda | \leftrightarrow \widetilde{\lambda}_{\dA}.
\end{equation}
We choose the representation of the Clifford algebra as
\begin{align}
    \sigma^\mu_{A \dA} &= (1, \sigma_z, \sigma_x, i \sigma_y) \\
    \widetilde{\sigma}^{\mu \dA A} &= \epsilon^{AB} \epsilon^{\dA \dB} \sigma^\mu_{B \dB}\\
    &= (1, -\sigma_z, -\sigma_x, -i \sigma_y).
\end{align}
With $p^\mu = (p^0, p^1, p^2, p^3)$, we define
\begin{align}
    p_{A \dA} &= p \cdot \sigma_{A \dA} = \begin{pmatrix} p^0 - p^1 & -p^2  + p^3 \\ -p^2 - p^3 & p^0 + p^1 \end{pmatrix} \\
    p^{\dA A} &= p \cdot \widetilde{\sigma}^{\dA A} = \begin{pmatrix} p^0 + p^1 & p^2  - p^3 \\ p^2 + p^3 & p^0 - p^1 \end{pmatrix}.
\end{align}
These satisfy the relation
\begin{equation}
    p^{\dA A} q_{A \dA} = 2 p \cdot q.
\end{equation}
The Lorentz generators acting on the spinor variables are defined as
\begin{align}
    (\sigma^{\mu \nu})_{A}^{\;\; B} &= \frac{1}{4} \sigma^\mu_{A \dC} \widetilde{\sigma}^{\nu \dC B} - \frac{1}{4} \sigma^\nu_{A \dC} \widetilde{\sigma}^{\mu \dC B} \\
    (\widetilde{\sigma}^{\mu \nu})^{\dA}_{\;\; \dB} &= \frac{1}{4} \widetilde{\sigma}^{\mu \dA C} \sigma^\nu_{ C \dB} - \frac{1}{4} \widetilde{\sigma}^{\nu \dA C} \sigma^{\mu}_{C \dB}
\end{align}
Note $\sigma^{\mu \nu}$ and $\widetilde{\sigma}^{\mu \nu}$ satisfy the self-duality and anti-self-duality properties
\begin{align}
    (\sigma_{\mu \nu})_{A B} &= \frac{1}{2}\epsilon_{\mu \nu \rho \sigma} (\sigma^{\rho \sigma})_{A B} \\
    (\widetilde{\sigma}_{\mu \nu})_{\dA \dB} &= -\frac{1}{2}\epsilon_{\mu \nu \rho \sigma} (\widetilde{\sigma}^{\rho \sigma})_{\dA \dB}.
\end{align}
The polarization vectors are defined by
\begin{equation}
    \varepsilon^{+}_{A\dA} = \sqrt{2} \frac{ \lambda_A \widetilde{\xi}_{\dA} }{ [ \lambda \xi] }\,, \hspace{1 cm}
    \varepsilon^{-}_{A \dA} =  \sqrt{2} \frac{\xi_A \widetilde{\lambda}_{\dA} }{ \langle \lambda \xi \rangle }.
\end{equation}
we have the useful identities
\begin{align}
    k_\mu \varepsilon^+_\nu (\sigma^{\mu \nu})_{AB} &= \frac{1}{\sqrt{2}} \lambda_A \lambda_B 
    \hspace{1 cm} k_\mu \varepsilon^-_\nu (\sigma^{\mu \nu})_{AB} = 0 \\
    k_\mu \varepsilon^-_\nu (\widetilde{\sigma}^{\mu \nu})_{\dA \dB} &= \frac{1}{\sqrt{2}} \widetilde{\lambda}_{\dA} \widetilde{\lambda}_{\dB} \hspace{1 cm} k_\mu \varepsilon^+_\nu (\widetilde{\sigma}^{\mu \nu})_{\dA \dB} = 0.
\end{align}
The Weyl tensor, defined to linear order in $G$ by
\begin{equation}
C_{\mu\nu\rho\sigma}(x)=\partial_{\rho}\partial_{[\mu}h_{\nu]\sigma}-\partial_{\sigma}\partial_{[\mu}h_{\nu]\rho}+\mathcal{O}(G^{2}).
\end{equation}
contains all of the gauge invariant information about the metric. Using \eqref{eq:resfr} this becomes

\begin{equation}
C_{\mu\nu\rho\sigma}^{\pm}(x)=\frac{16\pi G}{(2\pi)^{2}}\int d^2\lambda e^{-k\cdot x} k_{[\mu }\epsilon_{\nu]}^{\mp} k_{[\rho}\epsilon_{\sigma]}^{\mp}\mathcal{M}_{3}^{\pm}(k)\,,
\end{equation}
Because it is trace-free, it is equal to a self-dual part plus an anti-self-dual part. We can write them in spinor form using
\begin{align}
    C_{ABCD} &= \frac{1}{4}C^{-}_{\mu \nu \rho \sigma} (\sigma^{\mu \nu})_{AB} (\sigma^{\rho \sigma})_{CD} \\
    C_{\dA \dB \dC \dD} &= \frac{1}{4} C^{+}_{\mu \nu \rho \sigma} (\widetilde{\sigma}^{\mu \nu})_{\dA \dB} (\widetilde{\sigma}^{\rho \sigma})_{\dC \dD}.
\end{align}
then using the so-called Weyl decomposition we express the tensor as
\begin{equation}
C_{\mu\nu\rho\sigma}\sigma_{A\dot{A}}^{\mu}\sigma_{B\dot{B}}^{\mu}\sigma_{C\dot{C}}^{\mu}\sigma_{D\dot{D}}^{\sigma}=4(C_{ABCD}\epsilon_{\dot{A}\dot{B}}\epsilon_{\dot{C}\dot{D}}+\epsilon_{AB}\epsilon_{CD}C_{\dot{A}\dot{B}\dot{C}\dot{D}})\,.\label{eq:Cdecomp}
\end{equation}       
We shall now write down a spinorial expression for the Weyl Tensor. Let us define the spinors $\lambda^A$ and $\widetilde{\lambda}^{\dA}$ by
\begin{equation}
    k_{A \dA} = k_\mu \sigma^\mu_{A \dA} =  \lambda_{\dA} \widetilde{\lambda}_{A}.
\end{equation}
Notice that the above equation does not uniquely determine $\lambda^A$ and $\widetilde{\lambda}^{\dA}$ because one can always perform a rescaling $\lambda^A \to \alpha \lambda^A$, $\widetilde{\lambda}^{\dA} \to \widetilde{\lambda}^{\dA}/\alpha$ which corresponds to a $SO(1,1)$ little group transformation. To fix this redundancy, we will use the four-velocity of the stationary source as a reference vector. If $u^\mu = (1,0,0,0)$, $u_{A \dA}$ is simply given by
\begin{equation}
    u_{A \dA } = \sigma^0_{A \dA}.
\end{equation}
This allows us to relate $\lambda^A$ and $\widetilde{\lambda}^{\dA}$ by
\begin{equation}
    \widetilde{\lambda}^{\dA} = u^{\dA A} \lambda_A, \hspace{1 cm} | \lambda ] = u | \lambda \rangle
\end{equation}
Putting this all together we arrive at the final expressions 
\begin{align} \label{toprove1}
    C_{ABCD} &= G\int\frac{d^{2}\lambda}{2\pi}e^{-\frac{1}{2}\langle \lambda | x u | \lambda \rangle}\lambda_{A}\lambda_{B}\lambda_{C}\lambda_{D}\mathcal{M}_{3}^{-}\left(|\lambda\rangle,[\tilde{\lambda}|=\langle\lambda|u\right)\,, \\
    C_{\dot{A}\dot{B}\dot{C}\dot{D}}  &=  G\int\frac{d^{2}\tilde{\lambda}}{2\pi}e^{-\frac{1}{2}[ \lambda | x u | \lambda ]}\tilde{\lambda}_{\dot{A}}\tilde{\lambda}_{\dot{B}}\tilde{\lambda}_{\dot{C}}\tilde{\lambda}_{\dot{D}}\mathcal{M}_{3}^{+}\left(\langle\lambda|=[\tilde{\lambda}|\tilde{u},|\tilde{\lambda}]\right)\,. \label{toprove2}
\end{align}

\normalem
\bibliographystyle{jhep}
\bibliography{bhbib}

\end{document}